\documentclass[12pt,preprint,longnamesfirst]{aastex}
\usepackage{epstopdf}
\usepackage{amsmath}

\newcommand{\kms}{\ifmmode \mathrm{km~s^{-1}}\else km~s$^{-1}$\fi}
\newcommand{\smpy}{\ifmmode M_\sun~\mathrm{yr}^{-1}\else M$_\sun$~yr$^{-1}$\fi}
\newcommand{\lir}{\ifmmode L_\mathrm{IR}\else $L_\mathrm{IR}$\fi}
\newcommand{\lfir}{\ifmmode L_\mathrm{FIR}\else $L_\mathrm{FIR}$\fi}
\newcommand{\lsun}{\ifmmode L_\sun\else $L_\sun$\fi}
\newcommand{\msun}{\ifmmode M_\sun\else $M_\sun$\fi}
\newcommand{\rone}{$(\frac{\it{r}}{1~\rm{kpc}})$~}
\newcommand{\rfive}{$(\frac{\it{r}}{5~\rm{kpc}})$~}
\newcommand{\omone}{$(\frac{\Omega}{0.1})$~}
\newcommand{\nags}{\ion{Na}{1}}
\newcommand{\nad}{\ion{Na}{1}~D~}

\newcommand{\heo}{\ion{He}{1}~}
\newcommand{\mgb}{\ion{Mg}{1}~{\it b}}
\newcommand{\ooll}{[\ion{O}{1}]~$\lambda\lambda6300,~6364$}
\newcommand{\ot}{[\ion{O}{3}]}
\newcommand{\otl}{[\ion{O}{3}]~$\lambda5007$}
\newcommand{\stll}{[\ion{S}{2}]~$\lambda\lambda6716,~6731$}
\newcommand{\nt}{[\ion{N}{2}]}

\newcommand{\ntla}{[\ion{N}{2}]~$\lambda6548$}
\newcommand{\ntlb}{[\ion{N}{2}]~$\lambda6583$}
\newcommand{\ntll}{[\ion{N}{2}]~$\lambda\lambda$6548,~6583}
\newcommand{\cf}{\ifmmode C_f\else $C_f$\fi}
\newcommand{\co}{\ifmmode C_\Omega\else $C_\Omega$\fi}
\newcommand{\dvmax}{\ifmmode \Delta v_{max}\else $\Delta v_{max}$\fi}
\newcommand{\dvtau}{\ifmmode \Delta v_{maxN}\else $\Delta v_{maxN}$\fi}

\shorttitle{Neutral Gas Outflows and Inflows in Infrared-Faint
  Seyfert Galaxies}
\shortauthors{Krug, Rupke, \& Veilleux}
\begin{document}

\title{Neutral Gas Outflows and Inflows in Infrared-Faint Seyfert
  Galaxies}

\author{Hannah B. Krug\altaffilmark{1}, David
  S.~N. Rupke\altaffilmark{2}, \& Sylvain Veilleux\altaffilmark{1,3}}

\altaffiltext{1}{Department of Astronomy, University of Maryland, College Park,
  MD 20742; E-mail: hkrug@astro.umd.edu, veilleux@astro.umd.edu}
\altaffiltext{2}{Institute for Astronomy, University of Hawaii, 2680 Woodlawn
  Drive, Honolulu, HI 96822; E-mail: drupke@ifa.hawaii.edu}
\altaffiltext{3}{Also: Max-Planck-Institut f\"ur
    extraterrestrische Physik, Postfach 1312, D-85741 Garching,
    Germany}

\begin{abstract}

  Previous studies of the \nad interstellar absorption line doublet
  have shown that galactic winds occur in most galaxies with high
  infrared luminosities.  However, in infrared-bright composite
  systems where a starburst coexists with an active galactic nucleus
  (AGN), it is unclear whether the starburst, the AGN, or both are
  driving the outflows.  The present paper describes the results from
  a search for outflows in 35 infrared-faint Seyferts with 10$^{9.9}$ $<$ \lir/\lsun $<$
  10$^{11}$, or, equivalently, star formation rates (SFR) of $\sim$0.4 --
  9 \smpy, to attempt to isolate the source of the outflow.  We find
  that the outflow detection rates for the infrared-faint Seyfert 1s
  (6\%) and Seyfert 2s (18\%) are lower than previously reported for
  infrared-luminous Seyfert 1s (50\%) and Seyfert 2s (45\%). The
  outflow kinematics of infrared-faint and infrared-bright Seyfert 2
  galaxies resemble those of starburst galaxies, while the outflow
  velocities in Seyfert 1 galaxies are significantly larger.  Taken
  together, these results suggest that the AGN does not play a
  significant role in driving the outflows in most infrared-faint and
  infrared-bright systems, except the high-velocity outflows seen in
  Seyfert 1 galaxies.  Another striking result of this study is the
  high rate of detection of inflows in infrared-faint galaxies (39\%
  of Seyfert 1s, 35\% of Seyfert 2s), significantly larger than in
  infrared-luminous Seyferts (15\%).  This inflow may be contributing
  to the feeding of the AGN in these galaxies, and potentially
  provides more than enough material to power the observed nuclear
  activity over typical AGN lifetimes.

\end{abstract}


%%%%%%%%%%%%%%%%%%%%%%%%%%%%%%%%%%%%%%

\section{INTRODUCTION} \label{intro}

Galactic-scale gas outflows appear to play an important role in the
evolution of the universe, possibly providing an explanation for a
number of cosmological questions (\citealt{vcb05} and references
therein).  These galactic outflows are frequently observed in local
galaxies with high star formation rates and/or an active nucleus,
often extending on a scale of order of a few kiloparsecs or larger.

The high frequency of outflows measured in high-redshift, actively
star-forming objects such as Lyman break galaxies suggests that
outflows are a common stage in galactic evolution, with wind
velocities decreasing over time \citep{ss03,fr06}.  Such outflows may
be responsible for the deficit of baryons seen in our own Milky Way
galaxy and the mass-metallicity relation in external galaxies, if
winds are capable of preferentially ejecting metals into the
intergalactic medium (\citealt{l74,g02,thk04}).
The intergalactic medium can be heated as well as enriched with metals
by these galactic outflows, since energy inputs into the IGM on the
order of 10$^{56}$ ergs have been measured from galaxies with strong
outflows \citep{chk08}.  Additionally, outflows may quench star
formation by heating up cold gas and ejecting it from the host
\citep{b04,ssb05}.  Specifically, AGN-powered outflows have been proposed
as the cause of the drop in AGN luminosity at low redshift and the
cutoff at high luminosity of the galaxy luminosity function
\citep{sp99,c00,tsc06}.  These winds may also limit black hole and
spheroid growth, and be responsible for the observed tight correlation
between black hole mass and galactic spheroid mass \citep{fm00,md02,gzs04}.
Finally, winds driven by AGN and/or starbursts may help remove enough
nuclear material with low angular momentum early on in the evolution
of gas-rich systems to aid in the formation of large disks with high
specific angular momentum, more in line with the observations
\citep{bgs01,kk04}.

Studies performed on actively star-forming galaxies, near and far,
have revealed that outflow energetics increase with infrared (8-1000
$\mu m$) luminosity, \lir, or equivalently, star formation rate
(\citealt{m05,rvs05a}, hereafter RVS05a; \citealt{rvs05b}, hereafter RVS05b; \citealt{smnkl09,w09}).  Outflows
have also been detected in composite galaxies with both starbursts and
AGN \citep[e.g.,][]{is03,hs06}.  Seyfert 1 composites show
significantly higher outflow velocities than pure starbursts and
Seyfert 2 composites (\citealt{rvs05c}, hereafter RVS05c).  Hydrodynamical simulations of
starburst-driven outflows \citep[e.g.,][]{tsc06,cbsbh08} have
difficulties reproducing the very high velocities detected in Seyfert
1 composites.  The AGN in these objects thus appear to play a
significant role in driving these outflows, at least on small scales
\citep{ckg03}.  The situation in Seyfert 2 composites is more
ambiguous -- the AGN or starburst or both could be powering the
outflow in these systems \citep{sch85}.

The focus of this paper is on ``pure'' Seyfert galaxies with weak
infrared starbursts.  Our sample consists of 35 galaxies that are
faint in the infrared (10$^{9.9}$ $<$ \lir /\lsun~$<$~10$^{11.2}$),
equally split between Seyfert 1s and Seyfert 2s.  We compare the
results from our study with those from previous studies of starburst
and Seyfert ultraluminous infrared galaxies (ULIRGs; \lir/\lsun~$\ge$
10$^{12}$) and luminous infrared galaxies (LIRGs; 10$^{11}$ $<$
\lir/\lsun~$<$ 10$^{12}$) to isolate the role of the AGN in powering
these outflows.  As in previous studies, the IR-faint objects were
observed in the \nad $\lambda\lambda$5890,5896 doublet absorption
feature.  This Na feature has a low ionization potential (5.1 eV), so
it probes neutral gas, and it can be used to study the ISM due to its
high interstellar abundance.  All objects in the sample have $z$ $<$
0.05, and thus the \nad feature is found in the optical.  Velocity components in \nad absorption are
unambiguous indicators of outflowing (blueshifted velocities) or
inflowing (redshifted velocities) gas, as there must be a continuum
source behind the absorber.

The organization of this paper is as follows.  The sample is described
in Section \ref{sample}, including methods of selection and sample
properties in terms of redshifts, star formation rates, and spectral
types. The observations are discussed in Section \ref{obs}.  Section
\ref{analysis} describes the line fitting, the derivations of the
velocities and column densities, and estimates of the expected stellar
contributions to the measured \nad absorption.  In Section
\ref{outsec} we present the results on outflows, starting with the
Seyfert 2s and followed by the Seyfert 1s.  Next, Section \ref{insec}
describes the results on inflows in the same fashion as Section
\ref{outsec}.  This is followed by a discussion of the implications of
our findings in Section \ref{disc}, including comparisons with
previous studies, dynamical estimates, and possible connections
between inflows and nuclear structures.  We conclude in Section
\ref{conc} with a summary of our findings.

All calculations in this study assume present-day cosmological
parameter values, with $H_0$ $=$ 75 \kms~Mpc$^{-1}$, $\Omega_m$ $=$
0.3, and $\Omega_\Lambda$ $=$ 0.7.  All wavelengths quoted are vacuum
wavelengths.

%%%%%%%%%%%%%%%%%%%%%%%%%%%%%%%%%%%%%%%%%%%%%%

\section{SAMPLE} \label{sample}

The objects in the sample were selected using three main criteria:
They had to be optically classified as Seyferts and had to be faint in
the infrared (\lir~$\lesssim$ 10$^{11}$ \lsun).  All objects in the
sample were also selected to have $z$ $<$ 0.05 to focus on the
brightest sources and obtain reliable measurements.

Most galaxies in the sample are well-studied Seyfert galaxies with
extensive ancillary data \citep[e.g.][]{w92a,nw95}.  Roughly equal
numbers of Seyfert 1 and Seyfert 2 galaxies were chosen to allow
meaningful comparisons with the study of RVS05c on ULIRG
Seyferts.  The bolometric luminosities of the AGN in the present
sample are well matched to those in the composite systems of RVS05c.
This allows us to make direct comparisons between the two samples.
Basic properties for the galaxies in our sample can be found in
Table~\ref{objprop}.  The last column in that table lists the
references used for the compilation.

\subsection{Redshifts} \label{redsec}

Not all objects in this survey have well-constrained published
redshift measurements, so preference was given to values from stellar
absorption lines \citep{nw95,w03} when possible, followed by redshifts
taken from HI lines \citep{nw95,sob05}.  Redshifts measured from data
in the present survey were used if neither of the preceding were
available, but were given low priority due to systematic errors
($\sim$ 10$^{-4}$) associated with these measurements (e.g. it is
possible that part of the narrow line region (NLR) is outflowing so the
velocities based on the narrow emission lines are not necessarily
representative of the systemic velocity).  In this case, IRAF was used
to measure central wavelength of various emission lines, including
\ooll, \ntll, and \stll, and redshifts were calculated from those
values.

\subsection{Star Formation Rates} \label{sfr}

To estimate the star formation rates in these objects, we use their
far-infrared luminosity, $L$(40-120 $\mu$m), under the assumption
that the AGN does not contribute significantly to \lfir~\citep{s06}.
This is probably a good first-order approximation although AGN may
contribute $\sim$10-20\% to \lfir\ \citep{n07}.  \lfir~is calculated
using the prescription of Sanders \& Mirabel (1996), using data from
the {\em IRAS} Faint Source and {\em IRAS} Point Source Catalogs.
\lfir~is then used to calculate SFR using the relation in Kennicutt
(1998), $SFR = \lfir /(5.8 \times 10^9 \lsun)$ $M_\odot$ yr$^{-1}$. The
far-infrared luminosities and star formation rates of the objects in
our sample are listed in Table~\ref{objprop}.

\subsection{Spectral Types}

Spectral classifications were taken from the NASA Extragalactic
Database\footnote{\texttt{http://nedwww.ipac.caltech.edu/index.html}}
and confirmed through examination of emission lines in the spectra and
references in the literature.  Seyfert 1 -- 1.5 were labeled Seyfert
1s for the purpose of this study, while Seyfert 1.8 -- 2 were labeled
Seyfert 2s.  This is in keeping with the classification scheme of
RVS05c and allows for comparisons of our results on Seyfert 1 and
Seyfert 2 galaxies with those of previous studies.

\section{OBSERVATIONS} \label{obs}

All observations were taken over the course of three different
observing runs on the Kitt Peak 4-meter telescope.  Observing runs,
exposure times, and slit position angles for all objects are listed in
Table~\ref{objprop}.  All data were taken using the
Ritchey-Chr$\acute{\rm e}$tien Spectrograph with a
moderate-resolution grating, KPC-18C, in the first order along with
the GG-475 blocking filter.  The wavelength range was 1700 \AA,
allowing measurements of both the \nad absorption doublet and
H$\alpha$ $+$ \nt~ emission complex in the same exposure.  This
combination provides an average resolution of 85 \kms~with a
1$\farcs$25 slit.  Median signal-to-noise ratio near \nad was 85 per
\AA~and ranged from 23 to 337 per \AA, with seeing ranging from
$\sim$1 to 2\arcsec\ on average.

\section{ANALYSIS} \label{analysis}

\subsection{Line Fitting} \label{linefit}

Extraction and fitting procedures were performed as described in the
original study (RVS05a), but the basic procedure will be outlined
here; the method uses assumptions similar to those in a curve of
growth analysis but does explicitly fit the line profiles. The spectral extraction was done while keeping a constant
physical aperture size of $\sim$3 kpc; this aperture size was selected
to match the seeing and therefore reduce host galaxy contamination as
much as possible.  The reduction and calibration
were performed using standard IRAF reduction packages.  HeNeAr lamps
were used for wavelength calibration and stars such as G191-B2B and
BD+25 4655 were used as flux standards.

Once the spectra were reduced and calibrated, the region containing
the \nad absorption doublet, as well as the neighboring \heo
$\lambda$5876 emission line, was isolated.  A Levenberg-Marquardt
fitting routine was used to fit one to two components to the \heo
emission and the \nad absorption doublet line profile.  For emission lines, the formula used to fit
the intensity of the line was:
\begin{equation}
I(\lambda) = 1 + Ae^{((\lambda-\lambda_0)c/(\lambda_0b))^2}
\end{equation}
where the amplitude, A, the Doppler parameter, $b$ ($b$ =
$\sqrt{2}\sigma$), and the central wavelength, $\lambda_0$, were free
parameters.  For absorption lines, the
formula used to fit was:
\begin{equation}
I(\lambda) = 1 - C_f(\lambda) + C_f(\lambda)e^{-\tau_{1}(\lambda)-\tau_{2}(\lambda)},
\end{equation}
where $C_f$ is the covering fraction of the line (see
Section~\ref{mme}), and $\tau_{1}$ and $\tau_{2}$ represent the optical depths of each
of the two lines of the \nad doublet; $\tau_{1}$
corresponds to the line at 5896 \AA, and $\tau_{2}~=~2\tau_{1}$. 
In general, the optical depth $\tau_{i}$ is given by:
\begin{equation}
\tau_{i}(\lambda) = \tau_{i,0}e^{-(\lambda-\lambda_{i,0})^2/(\lambda_{i,0}b/c)^2}.
\end{equation}
Equations 2 and 3 were used to fit each line of the \nad
doublet.  For each \nad doublet pair, the covering fraction, $C_f$,
central optical depth and wavelength of the red line ($\tau_{1}$ and
$\lambda_{0,1}$), and line width, $b$, were free parameters in the fit.  
Covering fraction was assumed to be independent of wavelength (and the
same for both doublet lines), and the optical depths of the two
doublet lines are fixed by atomic physics (as stated above: $\tau_{2}$
$=$ 2$\tau_{1}$).  Because of these constraints, and the assumption of
a Gaussian in optical depth space, both $C_f$ and $\tau$ can be
constrained simultaneously for each doublet.  It should be noted,
however, that there is still some anticorrelation between $C_f$ and
$\tau$ in the fit due to uncertainty in line shape and relative
intensities.  In the case of multiple, Gaussian velocity components, 
each velocity component arises in both lines of the doublet, so that
$C_f$ and $\tau$ can be constrained independently for each velocity component.

\begin{figure}[t]
\epsscale{0.95}
\plotone{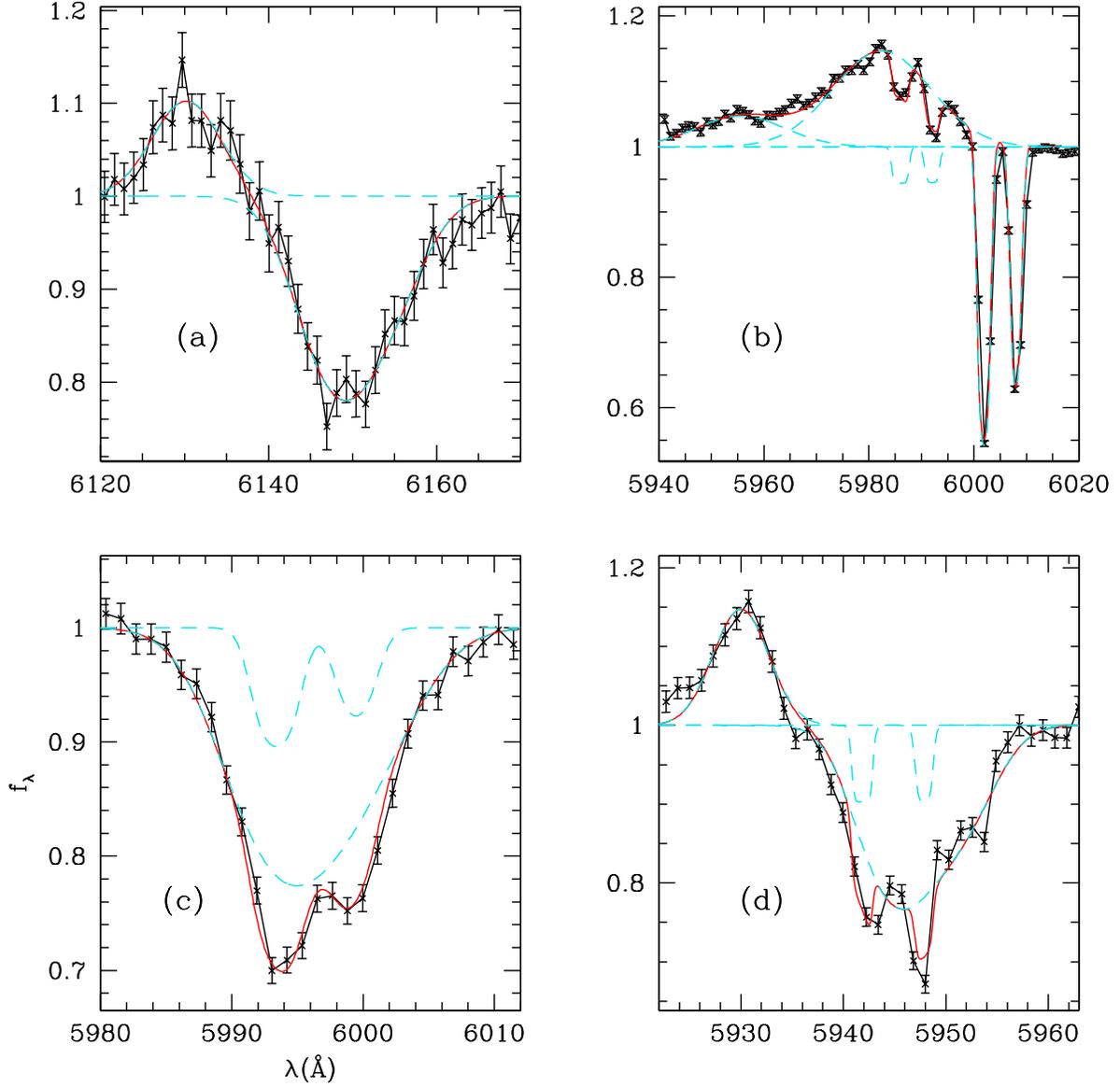}
\caption{Binned data showing the \heo and \nad complexes, plotted on an intensity versus
  wavelength scale, for four objects.  In all cases, blue dashed lines
  correspond to the individual \heo or \nad line fits, with solid red
  lines showing the combined \heo and \nad fit for the object.  Panels
  (a) and (b) show two Seyfert 1 galaxies, Mrk 506 and Mrk 6.  Mrk 506
  is fit with one \heo and one \nad velocity component.  Mrk 6 is fit with two \heo
  components and two \nad velocity components.  Panels (c) and (d) show two
  Seyfert 2 galaxies, Akn 79 and NGC 3786.  Akn 79 does not have a
  resolvable \heo line; two \nad components, one narrow and one
  broad, are used in the fit.  NGC 3786 is fit with one \heo
  component and two \nad, one narrow and one broad.  Note that the data for NGC
  3786 are low signal-to-noise compared to most other objects, and
  thus the lines are not quite as well constrained.}
\label{deblends}
\end{figure}

Initial guesses for these parameters were supplied to the fitting
program and a best fit was determined and plotted.  Either one or two
components were used to fit the \heo emission line.  Unless an
absorption fit with two separate velocity components (corresponding to
two separate \nad doublets) was considerably better than a fit with
one velocity component, only one velocity component (one \nad doublet)
was used to fit the absorption
features.  Only six objects showed convincing evidence for a
two velocity components in absorption.  Examples of the deblending of the
\nad feature for two Seyfert 1 and two Seyfert 2 objects can be
seen in Figure~\ref{deblends}.  After fitting was performed, a
Monte Carlo simulation with 1000 iterations was then run, in order to
determine the parameter errors.

\begin{figure}[t]
\epsscale{0.7}
\plotone{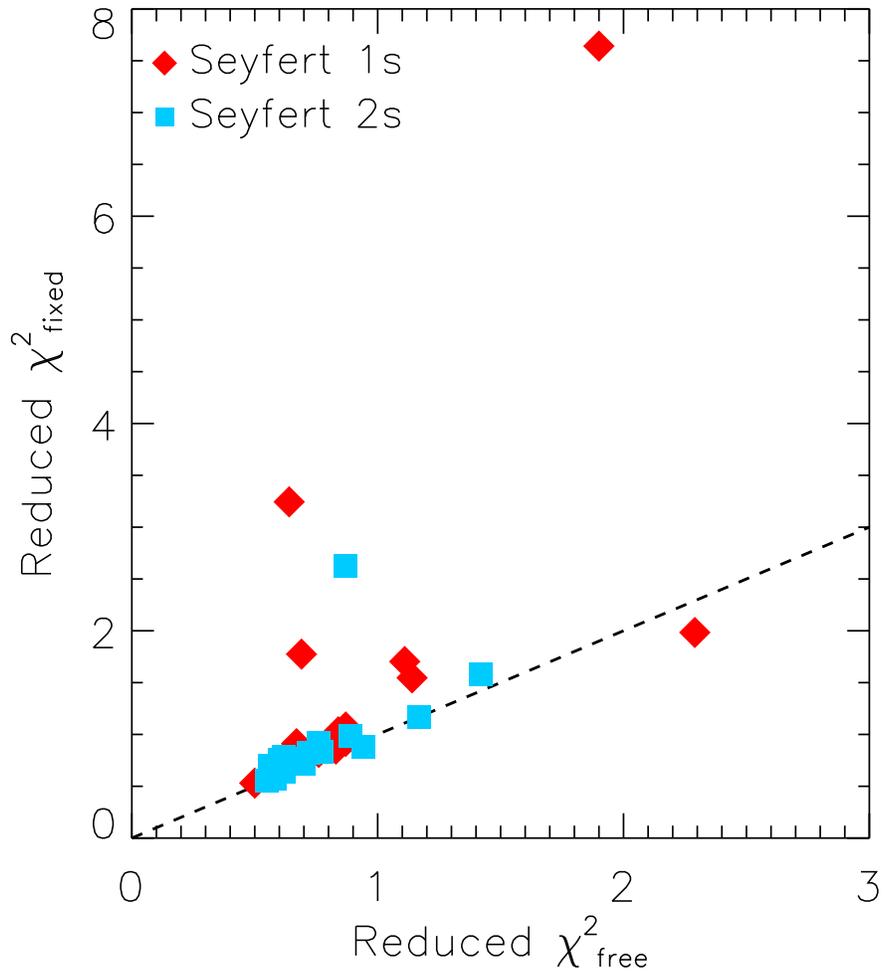}
\caption{Reduced $\chi^2$ values for free-floating and H$\alpha$-fixed
  \heo parameter fits. See Section \ref{linefit} for an explanation of
  these fits. The dashed black line corresponds to equal $\chi^2$
  values.  The free fits generally give lower $\chi^2$. The results
  from these fits were adopted in the present study.}
\label{redchi}
\end{figure}

Particular care is needed when fitting the \heo emission at 5876 \AA,
since the \heo emission line in these objects is generally quite broad
and has multiple components. So, for those objects showing \heo
emission lines, the H$\alpha$ $+$ \nt~ complex was also examined to
verify the accuracy of the fit to \heo. An interpolation script was
written in IDL to subtract out the \ntll~features and thus isolate the
H$\alpha$ line.  The resulting H$\alpha$ line was generally fit using
two components, one narrow and one broad.  The \heo $+$ \nad complex
was then refit, but this time parameters for \heo emission lines were
constrained to be those of the H$\alpha$ line.  The results from the
fits with fully fixed parameters (``fixed'') were then compared to
those from the fits with free-floating parameters (``free'');
agreement is observed within 2-5\%.  Again, the errors on these
parameters were determined from the Monte Carlo simulations after all
fits were completed. The reduced $\chi^2$ for the two types of fits
are compared in Figure~\ref{redchi}.  There are fewer points farther
away from a reduced $\chi^2$ of unity for the free-floating fit. We
therefore choose to use the results from the free-floating fits in the
following discussion (this is further justified in the next
section).

\subsection{Velocities} \label{velcalc}

\begin{figure}[t]
\epsscale{1.1}
\begin{center}
\plottwo{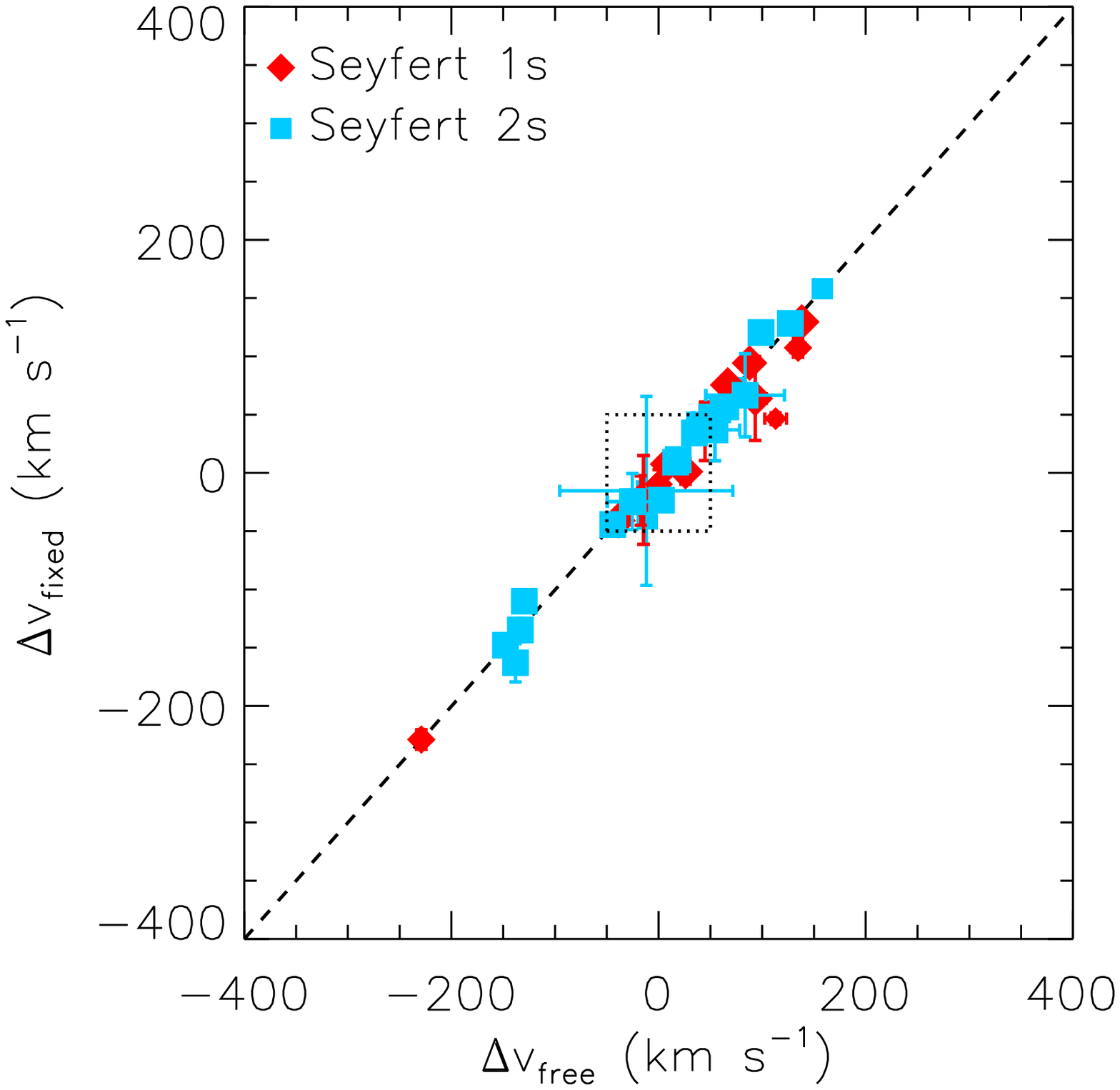}{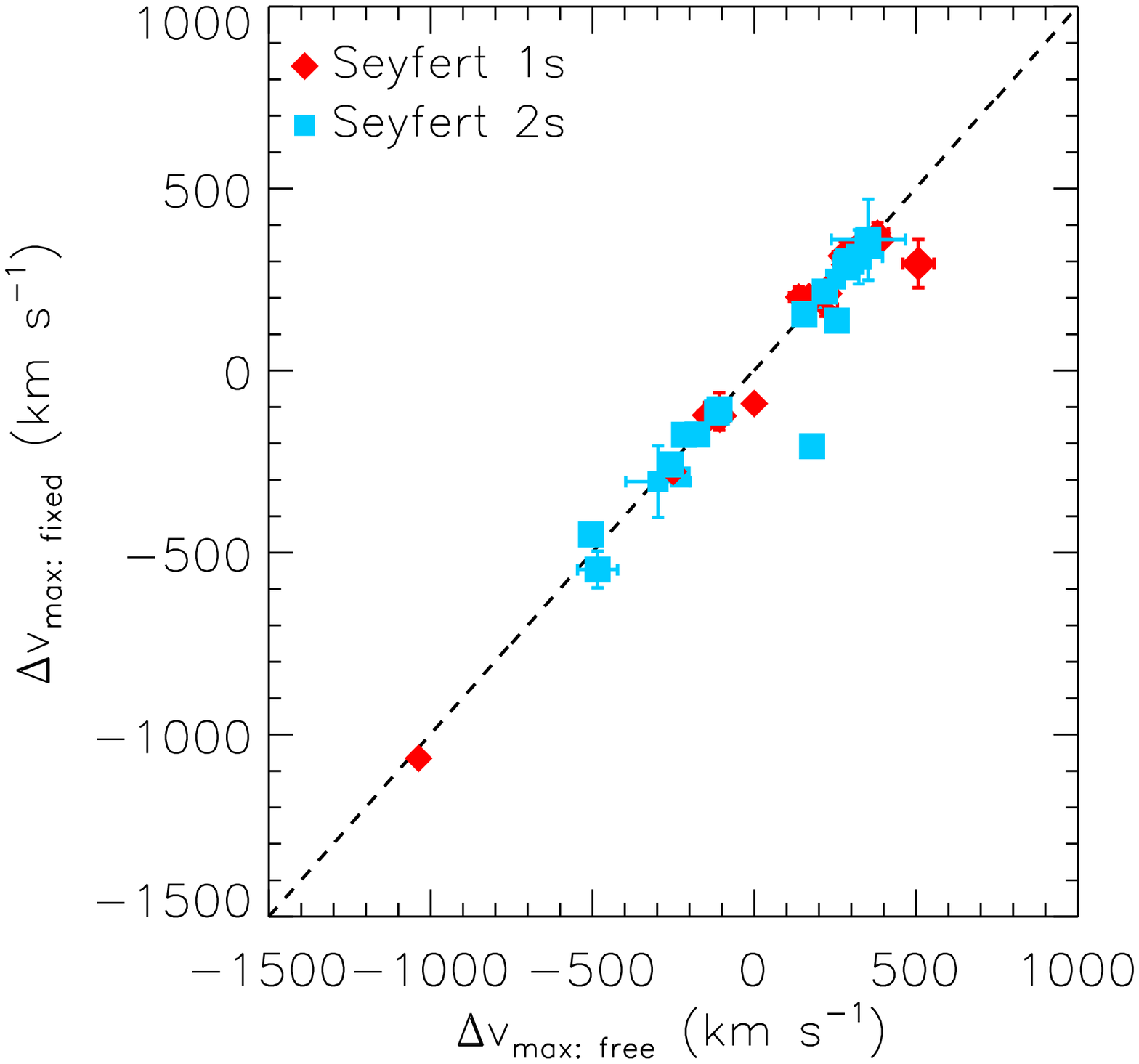}
\end{center}
\caption{Left: $\Delta v$ for H$\alpha$-fixed \heo fit versus $\Delta
  v$ for free-floating \heo fit.  Smaller points are objects with
  larger redshift uncertainty.  Dashed black line through center
  corresponds to equal fixed and free values, and all points outside
  dashed black box have $|\Delta v|$ $>$ 50 \kms.  Right: Same, but
  for \dvmax. The two types of fits give very similar results.}
\label{dvf1f4}
\end{figure}

Once fitting was done, velocities for these \nad absorption lines were
calculated, relative to the systemic velocities using redshifts as
outlined in Section \ref{redsec}.  In the following, $\Delta v$ is the
velocity difference between the \nad velocity centroid measured in the
fit and the systemic \nad velocity.  Errors were propagated through
both from the fit and the redshift measurement.  A ``maximum''
velocity of $\Delta v_{max} \equiv \Delta v$ - FWHM/2, which takes into
account both the shift and width of the absorption line, was also
computed.  Values of $\Delta v$ and $\Delta v_{max}$ were compared for
values resulting from the free \heo fit and the fixed \heo fit in
order to determine the impact of fixing the \heo parameters to those
of H$\alpha$. Plots of those comparisons are displayed in
Figure~\ref{dvf1f4}.  The agreement is good, with only three objects
showing values that are not equal within the uncertainties.  Of those,
only one has $|\Delta v_\mathrm{free} - \Delta v_\mathrm{fixed}|$ $>$
50 \kms, which would change its outflow or inflow classification.  In
all three of those outlying cases, the \heo line appears to be broader
than the H$\alpha$ line.  This is plausible, since all of these
objects are Seyfert 1 galaxies and so collisional and radiative
transfer effects in the broad line region may affect the \heo profile
in a different way than the H$\alpha$ profile.  As there is little
significant difference between the free-floating and fully-fixed fits,
our adoption of the free-floating fits in Section \ref{linefit} seems
appropriate.

After velocities were calculated, a detection criterion of $\Delta v$
$<$ $-$50 \kms~was used to determine whether or not an object possesses
an outflow.  Similarly, $\Delta v$ $>$ 50 \kms~was used as the cutoff
for the detection of an inflow.  A 2-$\sigma$ threshold in measurement
uncertainty ($|\Delta v|$ $>$ 2$\delta$($\Delta v$), where
$\delta$($\Delta v$) is the measurement error) was also required for
inflow/outflow classification, but this did not eliminate any objects
from being classified as showing inflow or outflow.  As outlined in
the initial study (RVS05a), the 50 \kms~cutoff accounts for potential
contamination from errors in wavelength calibration ($\lesssim$10
\kms), fitting ($\sim$10 \kms), redshift measurements ($\sim$
10-15 \kms), and the possibility of small blue- and redshifts due to gas in rotation.  Fit parameters and derived velocities are listed in
Table~\ref{compprop}.

\subsection{Column Densities} \label{cdcalc}

The optical depth $\tau$ and Doppler parameter $b$ as determined by the
fitting function were used to calculate column densities of \nad and H
along the line of sight.  As described in Spitzer (1978), the column
density of \nad in cm$^{-2}$ is given by:
\begin{equation}
\mathrm{\it{N}\rm{(Na~I~D)}} = \dfrac{\tau_{1,c} b}{1.497\times10^{-15}~\lambda_1 f_1},
\end{equation}
where $f_1$ is the oscillator strength, $\lambda_1$ is the rest frame
vacuum wavelength (\AA), and $\tau_{1,c}$ is the central optical depth
of the \nad $\lambda$5896 line.  The Doppler parameter, $b$, is in units
of \kms.  Values of $f_1$ $=$ 0.3180 and $\lambda_1$ $=$ 5897.55 \AA~
were taken from Morton (1991).

To properly calculate the column density of hydrogen, we must correct
for the effects of dust depletion and ionization.  This process is
also outlined in RVS05a, and uses empirical results that assume
Galactic depletion and an ionization fraction of 0.9.  The formula
used is:
\begin{equation}
\mathrm{\it{N}\rm{(H)}} = \mathrm{\it{N}\rm{(Na~I)}}(1-y)^{-1}10^{-(a+b)}
\end{equation}
Column densities are again given in cm$^{-2}$, $y$ is the ionization
fraction, and $a$ and $b$ are the Na abundance and depletion onto dust
in the object for which calculations are being performed,
respectively.  Whereas a near-IR luminosity-metallicity relation was
used to determine Na abundance in the previous RVS05b and RVS05c
studies, we here assume that these objects have solar Na
abundance. Calculated column densities for each object can be found in
Table~\ref{compprop}.

\subsection{Stellar \nad Contribution} \label{stelpop}

Stellar \nad absorption may contribute to the observed \nad feature.
Stellar features from other elements may also contaminate the
measurements if they are located close to the \nad feature.  The
original study used a scaling relation between \nad and
\mgb~equivalent width to determine the stellar \nad contribution, as
Na and Mg are created in a similar fashion (RVS05a).  However, the
spectral range for our data begins at roughly 5500 \AA~and so does not
include that \mgb~triplet, so we must use the other method outlined in
that study: stellar population synthesis models.  As in RVS05a, we
used the population synthesis code (Sed@.0
\footnote{\texttt{http://www.iaa.es/\textasciitilde rosa/ and
    http://www.iaa.es/\textasciitilde mcs/sed@}}) of Gonz\'{a}lez
Delgado et al. (2005), which combines a young (40 Myr), instantaneous
burst stellar population with an old (10 Gyr) population.  One model
uses a stellar mass ratio of 10\%/90\% for young {\em versus} old
populations, and the other uses 1\%/99\%.  In order to enhance weak
stellar features, the spectra were boxcar smoothed by ~150 \kms~and
convolved with a $\sigma$ $=$ 200 \kms~Gaussian.  Our spectra were
emission subtracted, using fitting parameters to remove \heo and leave
only the \nad absorption.  They were then overlaid with the stellar
population synthesis models.

\begin{figure}[t]
\epsscale{1.0}
\plotone{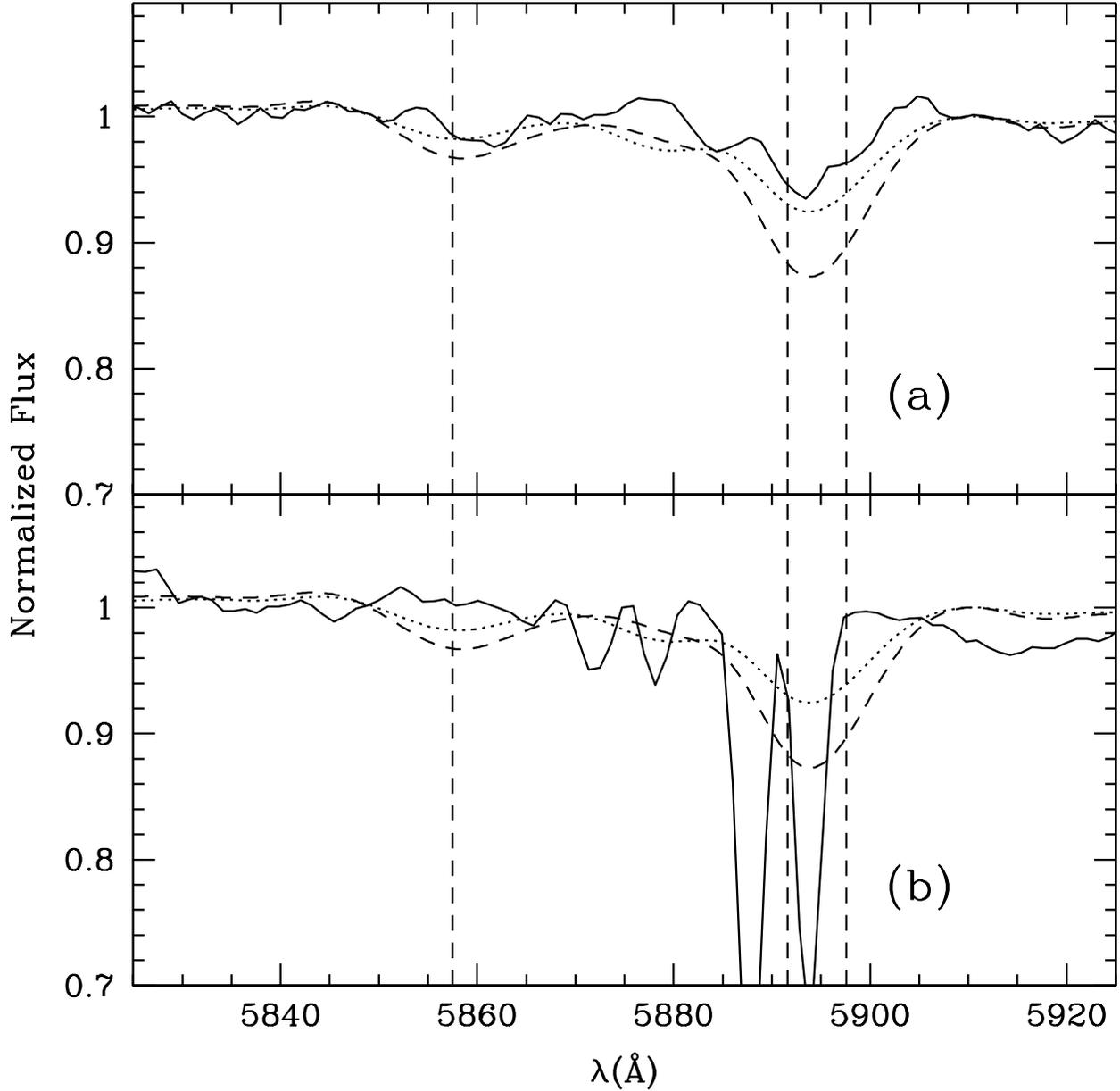}
\caption{Spectra of two Seyfert 1 galaxies, (a) Mrk~352 and (b) Mrk~6,
  in the \nad region. These spectra were He~I emission line
  subtracted, boxcar smoothed, and normalized to unity at 5910 \AA.
  The dotted line shows the 10/90 model (10\%/90\% stellar mass ratio
  for a young 40 Myr instantaneous burst population compared to an
  older 10 Gyr population) and the dashed line shows the 1/99 model
  (1\%/99\% young to old stellar mass ratio), as outlined in Section
  \ref{stelpop}.  Strong stellar features are marked with dashed
  vertical lines.  Mrk~352 shows absorption that is weaker than the
  stellar models, suggesting that the absorption is only stellar or
  atmospheric (the feature at $\sim$ 5880 \AA~was also determined to
  be atmospheric).  In contrast, Mrk~6 shows very strong absorption,
  which is consistent with predominant interstellar origin.}
\label{spop}
\end{figure}

Almost all objects with suspected outflows or inflows showed
absorption that was much deeper than the stellar absorption alone.
The extreme examples of Mrk~352 and Mrk~6 are shown in
Figure~\ref{spop}.  In Mrk~352, the results from stellar population synthesis show a
high likelihood that the bulk of the \nad absorption is stellar. In
contrast, Mrk~6 shows a definite high-velocity outflow.

\section{RESULTS: OUTFLOWS} \label{outsec}

\subsection{Seyfert 2} \label{outsyt}

Figure~\ref{sytspectrai} shows the \heo and \nad complexes for all of
the 17
Seyfert 2 galaxies in this study, plotted on a velocity scale based on
\nad $\lambda$5890 systemic.  Tables~\ref{outindiv} and \ref{avgpropout} list the
measured properties for each \nad outflow velocity component and overall
outflow averages, combined with data for Seyfert ULIRGs from RVS05c.

Out of a total of 17 Seyfert 2 galaxies, we detected blueshifted \nad
absorption with $\Delta v$ $<$ $-$50 \kms~in 3 objects, for an outflow
rate of 18 $\pm$ 9\%.  No objects were ruled out by the 2-$\sigma$
criterion in measurement uncertainty, which requires $|\Delta v|$ $>$
2$\delta$($\Delta v$) for a detection (see Section \ref{velcalc}).  Of
those objects which fit the outflow criteria, only one, NGC~7319,
showed convincing evidence for two velocity components in absorption.  The
implications of this low detection rate are discussed in Section
\ref{outdisc}.

Velocities for these objects were all within 20 \kms~of each other,
ranging from $-$130 to $-$148 \kms.  The maximum velocity, \dvmax~$\equiv$ $\Delta v$ $-$
FWHM/2, showed a much larger range (due to the
broader range of Doppler parameters), from $-$176 to $-$504 \kms,
with NGC~7319 showing the largest \dvmax~(Table~\ref{avgpropout}).
We further discuss these results in Section \ref{outdisc}. 

Previous studies have shown that it is common for Seyfert 2 galaxies
to have asymmetric emission lines with extended blue wings.  These
asymmetries are generally believed to be the result of outflowing gas
in the narrow line region of the AGN, and outflowing gas in the NLR
has been spatially resolved in both Seyfert 1 and Seyfert 2 galaxies
\citep{w68,wu87,ecp89,v91a,v91b,v91c,rck05}. This emission line asymmetry,
however, is not an unambiguous indicator of outflow in the way
absorption is since one cannot differentiate between outflowing
line-emitting material in front of the nucleus and infalling material
on the far side of the nucleus.  Blue-wing emission asymmetries can
only be interpreted as outflow if we are absolutely sure we are not
seeing the opposite side of the object \citep{ara06}.  The primary
focus of our paper is blueshifted \nad absorption, but it is of
interest to look for blue-wing emission line asymmetry in our Seyfert
2 galaxies to see if it is at all correlated with our findings on \nad
outflows.  In RVS05c, 75\% of Seyfert 2 nuclei showed blue
emission-line asymmetry (BELA), compared to a 45\% \nad outflow
detection rate.

We have used \ntll~to look for BELA in our objects (the
high-ionization \otl~line most commonly used in BELA detection is not
within our wavelength range).  The \ntll~ lines were separated from
H$\alpha$ by taking the blue wing of the \ntla~line and the red wing
of the \ntlb~line, and plotting them together about $\Delta v$ $=$ 0.
The two wings were then scaled appropriately to match each other in
intensity, and BELA determination was done by visual inspection.  5
out of 17 Seyfert 2 galaxies (29\%) showed evidence for BELA.  The most obvious case of BELA is NGC~5252.  Of those
which show BELA, only one (NGC~7319) has been determined to have an
\nad outflow, and one (Mrk~348) actually shows a \nad inflow (Section
\ref{insyt}), though it is the least asymmetric of all the BELA
objects.  Those five objects, along with a comparison object with a
\nad outflow but no BELA (Mrk~622), are shown in Figure~\ref{bela}.
Another object (NGC~3786; not shown in the figure) shows prominent
red-wing asymmetry; this object presents an \nad outflow rather than
an inflow.  Overall, we find no obvious correlation between emission
and absorption signatures of outflow, although the statistics are
poor.  Moreover, we cannot formally rule out the possibility that the
difference in outflow and BELA detection rates is simply due to our
use of \nt\ as a BELA probe since BELAs are more often seen in
high-ionization lines like \ot\ than in low-ionization lines like \nt\
\citep{v91a}.

\subsection{Seyfert 1} \label{outsyo} \label{outo}

Figure~\ref{syospectrai} shows the \heo and \nad complexes for all of
the 18 Seyfert 1 galaxies in this study, plotted on a velocity scale
relative to \nad $\lambda$5890 
systemic.  As for the Seyfert 2 galaxies, Tables~\ref{outindiv} and
\ref{avgpropout} list the measured properties for each \nad outflow
velocity component and overall outflow averages.

Only 1 of the 18 Seyfert 1 galaxies was found to have blueshifted
absorption with $\Delta v$ $<$ $-$50 \kms, and as with Seyfert 2
inflows, the 2-$\sigma$ criterion did not eliminate any objects
(Section \ref{velcalc}).  The outflow detection
rate is therefore only 6 $\pm$ 6\%.

Mrk~6 was the only Seyfert 1 galaxy determined to have an outflow.
Two velocity components were detected, each with a velocity higher
than those for the Seyfert 2 galaxies: one component with $-$229
\kms~and one with $-$1024 \kms, giving a \dvmax~= $-$1037
\kms~(Table~\ref{avgpropout}).  See Section \ref{outdisc} for a
discussion of these results.

\section{RESULTS: INFLOWS} \label{insec}

While only a few objects in our sample showed blueshifted \nad
absorption, a large percentage of objects in each Seyfert group showed
redshifted \nad absorption.  We consider this redshifted absorption to
be an unambiguous indicator of inflow, in the same way that
blueshifted absorption is an unambiguous outflow indicator, since the
continuum source (in this case the galaxy nucleus) must be behind the
absorber.  This section describes the results on inflow detection
rates and kinematics. Table~\ref{inindiv} lists the measured
properties of these inflows in individual objects. The implications of
these results are discussed in Section \ref{indisc}.

\subsection{Seyfert 2} \label{insyt}

The same general detection criteria used for determining outflows were
used for determining inflows: We required a redshifted absorption of
$\Delta v$ $>$ 50 \kms, along with the 2-$\sigma$ criterion (Section
\ref{velcalc}).  Out of the 17 Seyfert 2 galaxies, 6 showed \nad
inflow, for a detection rate of 35 $\pm$ 11\%.

Velocities for these six objects spanned the range from 51 to 127
\kms.  The maximum velocity, where in the case of inflows
\dvmax~$\equiv$ $\Delta v$ + FWHM/2, ranged from 155 to 352 \kms.  The
largest $\Delta v$ was measured for UGC 3995, although NGC~3362 has a
slightly larger \dvmax~than UGC 3995 due to the difference in Doppler
parameter.  The average properties of these inflows are listed in
Table~\ref{avgpropin} and discussed further in Section \ref{indisc}.

\subsection{Seyfert 1} \label{insyo}

Of the 18 Seyfert 1 galaxies, 7 showed \nad inflow, for a detection
rate of 39 $\pm$ 12\%, i.e.\ similar to that of the Seyfert 2s.
Again, no objects were ruled out by the 2$\sigma$ criterion
(Section \ref{velcalc}).

The Seyfert 1 inflow velocities spanned a similar range to those of
the Seyfert 2 velocities, from 67 to 138 \kms.  The Seyfert 1
\dvmax~values showed a broader range than that of Seyfert 2, from 169
to 507 \kms, and while Mrk~1126 has the highest $\Delta v$, NGC~7603
has the highest \dvmax.  Again, the average inflow properties are
listed in Table~\ref{avgpropin} and the results are discussed further
in Section \ref{indisc}.

\section{DISCUSSION} \label{disc}

\subsection{Outflows} \label{outdisc}

\subsubsection{Comparison with Previous Studies}

Since the purpose of this study is to determine whether starbursts or
AGN are the primary mechanism behind galactic outflows, it is
important to compare the results determined here to those of RVS05c
for the ULIRG Seyferts -- galaxies with both starburst and AGN. As
seen in Table~\ref{avgpropout}, the detection rates for the IR-faint
Seyfert 2s and Seyfert 1s are only 18\% $\pm$ 9\% and 6\% $\pm$ 6\%,
respectively, but rates were as high as 45\% $\pm$ 11\% and 50\% $\pm$
20\% for the IR-luminous Seyfert 2s and 1s (RVS05c).  Our detection
rates are even lower in comparison to non-Seyfert ULIRGs, since rates
of 80\% $\pm$ 7\% and 46\% $\pm$ 13\% were measured for low-$z$ and
high-$z$ ULIRGs in RVS05b, and a rate of 83\% was reported in Martin
(2005).  Poststarburst galaxies have also been found to have high
outflow detection rates on the order of 70\% \citep{tmd07}.  The
measured outflow detection rates in the IR-faint Seyfert galaxies are
therefore considerably smaller than those of pure starbursting
galaxies, poststarbursts, and AGN + starburst composites.

Previous observations have indicated that the outflow detection rate
increases with infrared luminosity (RVS05b, \citealt{smnkl09}).
Figure~\ref{detrateplots} shows the outflow detection rates of the
IR-faint Seyferts in our sample as well as the the IR-luminous
Seyferts and starburst ULIRGs and LIRGs of RVS05b and RVS05c as a
function of \lfir .  The increase in outflow detection rate with
\lfir~is apparent, despite the limited range in \lfir.  As \lfir~is
correlated with the star formation rate, this increase in the
detection rate with \lfir~suggests that star formation in all of
these systems is the main driver of neutral outflows detected in
\nags.  Geometry could also be playing a role, since AGN winds with
smaller opening angles (higher collimation) would reduce the detection
rate (see Section \ref{mme}). 

Histograms showing the distributions of all (negative and positive)
velocities in our current data as well as those of RVS05b and RVS05c
provide another point of comparison (Figure~\ref{velhistplots}).  The
left panel compares the velocities of IR-faint Seyfert galaxy
components from this study to the Seyfert and starburst ULIRGs \&
LIRGs of RVS05b and RVS05c.  We can safely rule out rotation as being a
primary cause for the motion of the \nad gas based on the fact that we
do not observe a symmetry about zero velocity in these histograms.
Were rotation the dominant gas motion, we should observe roughly equal
amounts of blueshifted and redshifted gas in each object and in the
overall distributions shown in Figure 8, assuming the \nad gas is
distributed more or less symmetrically around the center of rotation
and is unaffected by severe differential obscuration.  Negative velocities are found in a much
higher percentage of IR-luminous objects (Seyferts of RVS05c and
starbursts of RVS05b) than IR-faint objects, again indicating that
outflows are both stronger and more frequent on average in objects
with high SFR.  The right panel in Figure~\ref{velhistplots} combines
together the IR-faint and IR-luminous Seyferts from this study and
RVS05c and compares them to the starbursts of RVS05b.  In terms of the
negative velocities, the results for Seyfert 1 and Seyfert 2 galaxies
are quite different, with Seyfert 2 galaxies showing similar outflow
percentages to those of the ULIRGs \& LIRGs.  This suggests a physical
connection between the mechanisms that drive outflows in Seyfert 2s
and starbursting galaxies and a physical difference between the
mechanisms that cause outflows in Seyfert 1s and Seyfert 2s.
Interestingly, studies of {\em ionized} gas outflows have not found
such a dichotomy in velocities between Seyfert 1s and Seyfert 2s
(e.g., \citealt{rck01,vcb05,rck05,dck07}, and references therein).  This
suggests that the neutral gas probed by our \nad observations is not
kinematically related to the ionized material of the NLR.  This
difference is even more obvious when we also consider our results on
positive (inflow) velocities (Section \ref{incomp}).

Rigorous statistical tests confirm these results.  Kolmogorov-Smirnov
(K-S) and Kuiper tests were both used, as the K-S test has an inherent
bias in terms of differences in the mean and the Kuiper test does not.
Low values reported by both of these tests can help rule out two sets
of data having the same parent distribution. Results of these tests
are listed in Table~\ref{stattests}.  The small values of $P$(null)
for $\Delta v$ and \dvmax~when comparing IR-faint and IR-luminous
Seyferts suggest that they do not come from the same parent
distribution.  A more significant comparison comes from combining
together the results for the IR-faint Seyfert galaxies from this study
and those for the IR-luminous Seyfert galaxies from RVS05c, and
comparing them with those for the starburst ULIRGs and LIRGs from
RVS05b.  The same statistical tests were performed on these
distributions, first using all velocities, then the outflowing
components only.  The results are listed in Tables~\ref{stattests2a}
and \ref{stattests2b}, respectively.  The results in Table
\ref{stattests2a} indicate that Seyferts and starbursts do not share
the same parent velocity distributions; we return to this point in
Section \ref{indisc}.  When considering only the outflowing
components, all comparisons show low probability of originating from
the same parent distribution except when the Seyfert 2 galaxies are
compared with the starburst ULIRGs \& LIRGs.  In that case, both the
K-S and Kuiper tests return large $P$(null) values. This confirms
quantitatively that the outflows in these two classes of objects may
arise from the same physical process.

Figure~\ref{dvmaxplots} displays plots showing \dvmax~as a function of
far-IR luminosity (correlated with star formation rate) and of
galactic circular velocity (correlated with galactic mass) for
Seyferts from this study, IR-luminous Seyferts from RVS05c, starburst
LIRGs \& ULIRGs from RVS05b, and four starburst dwarf galaxies from
Schwartz \& Martin (2004).  The four dwarfs were added in order to see
if low-mass galaxies follow the same trends as high-mass systems.  The
IR-faint Seyfert 2 galaxies seem to follow the same trends as the
IR-bright galaxies, with \dvmax~increasing with both SFR and galactic
mass, while the outflow velocities of all Seyfert 1 galaxies lie above
these trends.  This again suggests a fundamental difference in the way
the winds in IR-faint/bright Seyfert 1 galaxies are powered compared
with the winds in IR-faint/bright Seyfert 2 and starburst galaxies.
We return to this issue in Section \ref{mme}.

\subsubsection{Outflow Dynamics} \label{mme}

Another useful comparison between IR-faint and IR-luminous Seyferts is
to look at the dynamical properties in addition to the kinematics.
Using calculated covering fraction, column density, and velocities, we
can estimate the mass, momentum, and kinetic energy of the neutral gas
phase of the ISM being probed by those winds.  We follow the method
outlined in the original study (RVS05b), which made the assumption
that the outflows are spherically symmetric mass-conserving free
winds, with a velocity and instantaneous mass outflow rate which do
not depend on radius within the wind and which are zero outside the
wind.  This method also assumed that the wind is a thin shell with a
uniform radius of 5 kpc. This value was based on actual spatial
measurements in some of the objects of RVS05c, but such spatial
measurements are not available for our IR-faint Seyfert galaxies.  In
the present study, we use a 5 kpc radius for both the Seyfert 2 and
Seyfert 1 galaxies.  This is different from the radius of 10 pc that
was used for the Seyfert 1 galaxies in the original study (RVS05c);
the radius in IR-faint Seyfert 1s was chosen to be the same as for
IR-faint Seyfert 2s to facilitate comparisons within the IR-faint
sample, and to that end, the IR-luminous Seyfert 1 dynamical values
from RVS05c have been scaled
up to a 5 kpc radius as well.  These results should be considered
order-of-magnitude estimates since they are based on a number of
largely unproven assumptions.  We have not listed all results here;
selected results can be seen in Figure~\ref{dmdtplots}.

For the objects in the present study, we assume a modest value of 0.3
for the large-scale opening angle (\co), which is the typical value
for local disk winds \citep{vcb05}.  Following the method of RVS05b, the covering fraction is used to
parameterize the clumpiness of the wind, or it may reflect the global
solid angle subtended by the wind when viewed from the galactic center.  Using the value of
$\langle$\cf$\rangle$ as listed in Table~\ref{avgpropout}, this yields
a global covering factor of $\Omega$ $\sim$ 0.1 for both Seyfert 1s and
Seyfert 2s.  These values are not well constrained, especially for the
Seyfert 1s since only one outflow was detected in this type of
Seyfert.  The $\Omega$ value is inconsistent with values of $\Omega$
$\sim$ 0.5-1.0 calculated by Crenshaw et al.~(2003b) from UV absorption lines in
local Seyfert 1s.  This low $\Omega$ value could imply that the winds
we are seeing are collimated rather than wide-angle, and thus our lack
of outflow detection in some of these objects may be due to their
orientation relative to us rather than a complete lack of neutral gas
outflow.  The influence of host galaxy inclination on column density
was explored using extinction-corrected axis ratios from the de
Vaucouleurs Third Reference Catalog \citep{rc3}, but no conclusion
could be drawn due to small-number statistics.  Additionally, these
$\Omega$ values will be low in comparison to
values from other studies; there is contribution here from the
background galaxy, and many of the outflows we have found are small-scale. Comparisons with the results of UV absorption line studies
should be considered with caution, though, since UV absorbers are typically of
much higher ionization (\ion{N}{5}, \ion{C}{4}) and likely located much closer to the
AGN than \nad absorbing material.  The \nad absorbers may also be
affected more strongly by host galaxy contamination than the UV
absorbers.  Blueshifted absorption lines have also been detected in
X-ray spectra for a number of Seyfert galaxies, including Mrk 6, with column densities
of significantly higher order of magnitude than we have found here
for \nad in the optical (e.g., \citealt{mew95,fbe99,mmwe01,knb03,vu08}).  Again, we
caution comparing these results with our own, as it has been noted
that these X-ray lines are much higher ionization (\ion{O}{7},
\ion{O}{8}) and are often intrinsically related to the aforementioned UV absorption lines.

One particularly interesting dynamical quantity is the mass outflow
rate, since galactic outflows may contribute to the IGM enrichment and
are possibly a quencher of star formation \citep{tmd07}.  Plots of
$dM/dt$ for the IR-faint Seyferts, IR-luminous Seyferts, and
starbursting LIRGs and ULIRGS from RVS05b and RVS05c, and the dwarfs
of Schwartz \& Martin (all calculations are based on a absorber radius
of 5 kpc) are presented in Figure~\ref{dmdtplots}.  We see a general trend of mass outflow rate
increasing with both \lfir~and galactic mass.  There is a considerable
difference in the mean $dM/dt$ rates between IR-luminous Seyfert 1s and IR-luminous Seyfert 2s.  The
momentum and energy, as well as outflow rates for those quantities,
are also significantly higher for IR-luminous Seyfert 1s than IR-luminous
Seyfert 2s and all IR-faint Seyferts.  The overall good
agreement in detection rate and kinematics between IR-faint
and IR-bright Seyfert 2s and starburst ULIRGs \& LIRGs
(Figures~\ref{detrateplots}-\ref{dmdtplots}) suggests that the
outflows in all of these objects are powered by star formation, while
the marked differences when IR-faint Seyfert 1 galaxies are considered
suggest that the AGN plays an important role in driving the (generally
high velocity) outflows in Seyfert 1 galaxies.  Escape fractions were not calculated here since too few
objects have outflow velocity components.

We can also look at energy outflow rates in order to determine what
role these outflows could play in galactic feedback.  The rates that
we have calculated can be found in Table~\ref{outindiv}, though it
must be cautioned that these numbers are a function of our uncertain $\Omega$
value, as well as our assumed absorber radius of 5 kpc.  For the Seyfert 2 outflows, the average
energy outflow rate was found to be $\sim$ 10$^{41.1}$ \omone \rfive ergs
s$^{-1}$, or $\sim$ 10$^{7}$ \omone \rfive \lsun.  In comparison to the
average bolometric luminosity for these objects ($\sim$ 10$^{10}$
\lsun, taken from \citealt{wu02}), the energy
outflow rates are only $\sim$1\% of the host galaxy luminosity.  This
indicates that outflow energetics in our Seyfert 2 galaxies are not
strong enough to play a large role in galactic feedback.  The findings
of Schlesinger et al.~(2009) for Mrk 573 are in agreement with our conclusion.
For the only Seyfert 1 galaxy that we
have found to have an
outflow, Mrk 6, the energy outflow rate of its lower velocity
component (10$^{41.6}$ \omone \rfive ergs s$^{-1}$) is again $\sim$1\% of
the host bolometric luminosity.  This indicates that this Seyfert 1 outflow
is also not energetic enough to play a vital role in galactic feedback,
and thus previous findings are consistent with our results \citep{kne07,smkn09}.  If we look at the higher velocity outflow found
in Mrk 6, its energy outflow rate (10$^{42.9}$ \omone \rfive ergs s$^{-1}$)
is $\sim$5\% of the host bolometric luminosity, which is higher than
for the Seyfert 2 galaxies but again likely not strong enough to
influence the evolution of its environment.  However, one should note that these values
are all highly dependent on the calculated value of $\Omega$ and
assumed value of $r$ and are thus uncertain in comparison to our measured velocities. 

\subsection{Inflows} \label{indisc}

\subsubsection{Comparison with Previous Studies} \label{incomp}

An unexpected result of the present study is the high detection rate
of inflows in IR-faint Seyferts (39\% $\pm$ 12\% for Seyfert 1s, 35\%
$\pm$ 11\% for Seyfert 2s). In contrast, only $\sim$15\% of the
IR-luminous objects in RVS05b and RVS05c showed redshifted \nad
absorption.  This difference is clearly seen in the left panel of
Figure~\ref{velhistplots}. Interestingly, a recent search for outflows
in the AEGIS database has also revealed an excess of inflows among
AGN-powered systems \citep{smnkl09}.  Inflow has been observed in at
least one Seyfert 1 galaxy, NGC 5548, using ionized gas detected in
the UV, though detections have not been reported for such a large
number of objects as 
we have found here \citep{mew99}.  There has been one tentative observation
of redshifted X-ray absorption in a Seyfert 1 galaxy, but the authors caution
that the significance of the absorption line they have measured is
highly uncertain \citep{dcm05}.

When we examine the inflow data on the IR-faint galaxies alone, we
find no significant correlation between \dvmax~and galactic mass (left
panel in Figure~\ref{inplots}).  Neither is there any obvious trend
with Seyfert type or far-infrared luminosity (right panel of
Figure~\ref{inplots}), in contrast to the trends seen for the
outflowing gas.  Additionally, we see inflows in nearly the same
fraction of Seyfert 1s and Seyfert 2s, indicating no particular trend
with Seyfert type.  As for the detected outflows, the influence of host galaxy inclination on column density
was explored using extinction-corrected axis ratios \citep{rc3}, but
again, no conclusion could be drawn due to small-number statistics.  

Next, the velocity distributions of the inflowing components for the
IR-faint and IR-luminous Seyfert 1 galaxies were combined together and
compared with the combined distribution of IR-faint and
IR-luminous Seyfert 2 galaxies. K-S and Kuiper tests were performed on
$\Delta v$, \dvmax, and Doppler parameter, following the same
procedure as in Section \ref{outdisc}, and the results are listed in
Table~\ref{stattests2c}.  They confirm the lack of obvious differences
in the inflow properties between the two types of Seyfert galaxies.

\subsubsection{Search for Connection between Inflows and Nuclear
  Structures} \label{nuc}

The \nags\ absorption infall velocities often extend to relatively high
values so we favor a nuclear location for this gas rather than a
galactic origin (e.g. galactic fountains, \citealt{fr06}). Nuclear
accretion of cool gas like \nags\ with T~$\sim$~100 K can provide fuel
not only for star formation but also for nuclear activity
\citep{smo08}.  Various mechanisms have been proposed to help reduce
the large angular momentum of the gas in the nuclei of galaxies. These
include nuclear ($\lesssim$ 1 kpc) bars and spirals
\citep{pm02,m03a,m03b,dm09} and gravitational interactions with
neighboring galaxies \citep{ckga03}.  Nuclear bars are thought to form
when a large galactic bar forces gas inwards, creating a gaseous disk,
and instability causes formation of a small gas bar near the nucleus
(\citealt{s89,mp01}).  Stellar bars could also accomplish the same
thing.  Both bar types are capable of removing angular momentum from
gas rotating near the nucleus \citep{mp01}.  Nuclear spirals have been
proposed as another AGN fueling mechanism since shock fronts that
occur at their boundaries can take away angular momentum from local
gas and cause material to fall in towards the black hole \citep{m03b}.
However, these nuclear structures do not necessarily lead to AGN
fueling since they are present in a equally large fraction of
non-active galaxies \citep{pm02}.  Star formation may occur in these
objects and disrupt AGN-fueling inflows \citep{dm09}.

We revisit this issue here by looking for the presence of nuclear
spirals or bars in the IR-faint Seyfert galaxies with inflow
signature.  We have compared our results to nuclear structure surveys
done by three different groups \citep{mgt98,m03a,dck06}.  Nuclear dust
structures in objects in common with these studies are classified into
five distinct morphological categories (see Column (11) of
Table~\ref{objprop}): irregular dust, dust filaments, nuclear dust
ring, nuclear dust bar, and nuclear dust spiral.  Of the thirteen
objects which show inflows, five show evidence for nuclear dust
spirals, bars, or rings, one shows evidence for dust filaments, one
for irregular dust, and six show no sign of nuclear dust structure.
The significant fraction of objects that show both inflows and nuclear
dust spirals, bars, rings, and/or filaments lends credence to the idea
of a connection between morphology and kinematics.  Of the seven
objects with irregular dust or no dust, five have nearby companion
galaxies, so tidal forces due to interactions with these companions
may cause AGN fueling \citep{rvb95,hc99,ssh07}.  This leaves only two
objects with inflow which do not show nuclear structure or a
companion: Akn~202 and Mrk~1018.  There are also two objects in our
study with nuclear dust structure that show measurable {\em outflow}
rather than inflow (NGC~3786 and NGC~7319), and thus whether we
measure outflow rather than inflow may be a consequence of our line of
sight to the nucleus (inflow and outflow may be occurring in different
planes, or in the same plane but over a different range of azimuthal
angles), rather than a lack of inflow in the object.  Measurements of
the line-of-sight velocity field with a resolution of $\sim$10s of pc
will be needed to disentangle the geometry of the inflows/outflows
detected in our data (e.g., the study of NGC~1097 by \citealt{dm09, vdv09}).

\subsubsection{Inflow Dynamics} \label{mmein}

The same method used to calculate mass, momentum, and kinetic energy
for outflows (Section \ref{mme}) was used for inflows, but the
characteristic absorber radius was reduced to 1 kpc, a rough upper
limit to the scale of the circumnuclear structures (nuclear
bars/spirals) believed to be responsible for feeding the AGN (Section
\ref{nuc}, \citealt{m03a}).  Again, these dynamical quantities are
rather uncertain, but are calculated to find out at least roughly how
these inflows compare with the mass accretion rates necessary to power
the AGN in these systems.  Table~\ref{inindiv} lists the mass,
momentum, and kinetic energy calculated for all objects with measured
inflows.  The mass accretion rate for these inflows ranges from just
under 1 \rone \smpy~to just under 5 \rone \smpy.  For comparison, the
mass accretion rate needed to power an AGN is $\dot{M} =
L_{\mathrm{bol}} / c^2 \eta,$ where $L_{bol}$ is the bolometric
luminosity of the AGN, $\dot{M}$ is the mass accretion rate, $c$ is
the speed of light, and $\eta$ is an efficiency factor dictating how
much of the rest mass of the material being accreted is turned into
radiation.  If we take $\eta$ to be $\approx$ 0.1 \citep{rif08} and
allow the bolometric luminosity to be $\sim$ 10$^{44}$ ergs s$^{-1}$,
typical for Seyfert galaxies \citep{pr88, ckg03}, then we find that a
mass accretion rate of $\dot{M}$~$\sim$~10$^{-2}$ \smpy~is required.
Even with our rough order of magnitude estimates, the mass accretion
rates of all of our observed \nad inflows are well above the amount
necessary to power the AGN.  Thus the inflows that we are measuring
carry enough material to fuel the AGN in these objects, even if only
$\sim$1\% of this material makes its way down to the AGN.  The total
infalling mass of $\sim$~10$^{7}$ \rone \msun, estimated from our
data, is enough to sustain nuclear activity over typical AGN lifetimes
($\sim 10^7 - 10^8$ yrs; \citealt{mt02,cr04}).

\section{SUMMARY} \label{conc}

The main results from our study of \nad absorption in infrared-faint
Seyfert 1 and Seyfert 2 galaxies can be summarized as follows:

\begin{itemize}
\item {\em Outflow Detection Rates and Kinematics:} The rates of
  detection of outflows in IR-faint Seyfert 1 and 2 galaxies are lower
  than previously found in IR-bright Seyferts.  Outflows were found in
  $\sim$18\% of IR-faint Seyfert 2s in our sample, compared with
  $\sim$45\% among the Seyfert 2 ULIRGs of RVS05c.  Only one out of 18
  Seyfert 1 galaxies in our sample shows evidence for a wind, far
  lower than the $\sim$50\% reported for Seyfert 1 ULIRGs in RVS05c.
  Interestingly, the outflow detection rate and velocities of IR-faint
  Seyferts follow the same trends with infrared luminosity and galaxy
  mass as IR-bright systems.  The outflow kinematics of Seyfert 2
  galaxies resemble those of starburst galaxies, while the outflow
  velocities in Seyfert 1 galaxies are significantly larger.  These
  results suggest that the AGN is contributing to the neutral outflows
  in Seyfert 1 systems, while the starburst is the main driver of the
  outflows in all Seyfert 2 galaxies.  Differences in wind angular
  extent (e.g. AGN-driven outflows in Seyfert 1s are more collimated
  than starburst-driven outflows) may also explain some of these
  results.
    
\item {\em Outflow Dynamical Estimates:} The mass, momentum, and
  kinetic energy of the material involved in these outflows were
  estimated assuming a constant characteristic radius of 5 kpc.  The
  dynamical properties of the outflows in Seyfert 2 galaxies are
  similar to those of the starburst ULIRGs \& LIRGs, but differ
  significantly from those of the Seyfert 1s.  This again suggests a
  fundamental physical difference between the outflows of Seyfert 1s
  and those in the other objects.  An attempt to determine the influence of
  host galaxy inclination on these outflows was inconclusive.
  Measured energetic rates do not appear large enough to play a
  significant role in galactic feedback, but these values are uncertain.  

\item {\em Inflow Detection Rates and Kinematics:} A striking result
  of this study is the high rate of detection of spatially-unresolved
  redshifted \nad absorption, which we interpret as nuclear inflows
  (39\% and 35\% inflow detection rates for Seyfert 2s and 1s), with
  maximum velocities (321 and 291 \kms~on average).  This is evidence
  for the existence of some mechanism capable of removing angular
  momentum from the circumnuclear gas in these objects.  Nuclear bars
  and spirals, as well as interactions with nearby galaxies, may play
  a role in this process.

\item {\em Inflow Dynamical Estimates:} Mass, momentum, and kinetic
  energy were estimated for the inflows, using a characteristic radius
  of 1 kpc, consistent with the observed sizes of nuclear bars and
  spirals in these systems.  While these estimates are uncertain, we
  find that the total infalling mass and infalling mass rates are more
  than enough to power the AGN in these systems for typical AGN
  lifetimes.  As with the outflows, an attempt was made to determine
  the influence of inclination on column density, but no conclusion could be drawn.
\end{itemize}

\clearpage

\acknowledgements

H.B.K., D.S.N.R., and S.V.\ were supported in part by NSF through
contract AST/EXC 0606932.  S.V.\ also acknowledges support from a
Senior Award from the Alexander von Humboldt Foundation and thanks the
host institution, MPE Garching, where some of this work was performed.
H.B.K.\ would like to acknowledge Stacy McGaugh and Massimo Ricotti
for helpful discussions during the writing process.  The
authors would also like to thank Michael McDonald for his assistance
with Perl computation code.  The observations reported here were
obtained at the Kitt Peak National Observatory, National Optical
Astronomy Observatory, which is operated by the Association of
Universities for Research in Astronomy (AURA), Inc., under cooperative
agreement with the National Science Foundation.  This research has
made use of the NASA/IPAC Extragalactic Database (NED), which is
operated by JPL/Caltech, under contract with NASA.

%%%%%%%%%%%%%%%%%%%%%%%%%%%%%%%%%%%%%%

%%%%%%%%%%
% TABLES %
%%%%%%%%%%

\clearpage

\begin{deluxetable}{llccrcclrccc}
%\rotate
\tabletypesize{\footnotesize}
\tablecaption{Galaxy properties and observing logs\label{objprop}}
\tablewidth{0pt}
\tablehead{
\colhead{Name} & \colhead{$z$} & \colhead{\lir} & \colhead{\lfir} &
\colhead{SFR} & \colhead{$v_c$} & \colhead{$W_{eq}$} & \colhead{Run} & \colhead{$t_{exp}$} & \colhead{PA}  & \colhead{Structure} & \colhead{Refs} \\
\colhead{(1)} & \colhead{(2)} & \colhead{(3)} & \colhead{(4)} &
\colhead{(5)} & \colhead{(6)} & \colhead{(7)} & \colhead{(8)} & \colhead{(9)} & \colhead{(10)} & \colhead{(11)} & \colhead{(12)}
}
\startdata
\multicolumn{11}{c}{\bf{Seyfert 2s}}\\ 
\tableline
Akn 79	&   0.01743 &   \nodata &   \nodata &  \nodata &	231 &	3.43 &  2005 Sep 05 &  4500 &     0 &	- & 1\\
Arp 107A &  0.03463 &   10.63 &	    10.09 &     2.14 &	308 &	2.20 &	2006 Apr 02 &  5400 &     0 &	- & 1\\
Mrk 348	&   0.01516 &	10.40 &	    9.89 &      1.33	&       191 &	1.53 &	2006 Nov 23 &  4800 &     0 &	- & 1\\
Mrk 622	&   0.02347 &	10.72 &	    10.26 &     3.15 &	301 &	2.03 &	2006 Apr 02 &  5400 &	0 &	I & 1,4,7\\
Mrk 686	&   0.01420 &	10.04 &	    9.64 &      0.76 &	558 &	2.98 &	2006 Mar 31 &  4500 &	90 &    S & 1,5,6\\
Mrk 1157 &  0.01510 &	10.41 &	    10.13 &     2.32 &	585 &	1.99 &	2005 Sep 04 &  3540 &	0 &	R & 1,4,7\\
NGC 1358 &  0.01351 &	10.35 &	    9.37 &      0.41 &	299 &	3.73 &	2006 Nov 21 &  4800 &	0 &	- & 1\\
NGC 1667 &  0.01517$^\mathrm{a}$ &	10.97 &	    10.66 &     7.84 &	580 &	3.99 &	2006 Nov 21 &  4800 &	0 &	S & 1,4,6\\
NGC 3362 &  0.02767 &  	\nodata &   \nodata &	\nodata &       358 &	1.92 &	2006 Mar 31 &  4500 &	90 &	F & \ 1,7\\
NGC 3786 &  0.00903 &	\nodata	&   \nodata &   \nodata	& 367 &	3.11 &
2006 Apr 02 &	5400	& 22	&	R,S & 1,6,8\\ 
NGC 4388 &  0.00839 &	10.41 &	    10.34 &     3.78 &  414 &	2.27	&	2006 Apr 02	&	5400	& 0	&	I & 1,5,6\\
NGC 5252 &  0.02308 &	10.39 &	    9.84 &      1.18 &  340 &	3.72 &	2006 Apr 02	&	5400	& 0	&	R,S & 1,6,7\\
NGC 5728 &  0.00932$^\mathrm{a}$ &	10.60 &	    10.32 &     3.63	& 434
&	2.25	&	2006 Apr 02	&	3600	& 0	&	- & 2,4\\
NGC 7319 &  0.02236 &	10.56 &	    10.21 &     2.83	& 210
&	6.61	&	2005 Sep 08	&	3240	& 0	&	F & 1,7\\
NGC 7672 &  0.01348 &	9.91 &	    9.64 &      0.76	& 363
&	1.39	&	2006 Nov 21	&	7200	& 0	&	- & 1,4\\
NGC 7682 &  0.01707 &	\nodata	&   \nodata &   \nodata	& 412
&	2.06	&	2006 Nov 21	&	6000	& 0	&	S & 1,6\\
UGC 3995 &  0.01575 &	10.32 &	    9.76 &      0.99	& 250
&	3.25	&	2006 Mar 31	&	4800	& 90	&	- & 1,4\\
\tableline 
\multicolumn{11}{c}{\bf{Seyfert 1s}}\\ 
\tableline
Akn 202	&   0.02872$^\mathrm{b}$	&	10.71	&  10.15 &	2.43	&	610
&	0.44	&	2006 Nov 23	&	6000	&	0
&	- & 4	\\
Akn 564	&   0.02468$^\mathrm{a}$ &	10.75 &	    10.14 &     2.40 &	443 &	0.00 &  2005 Sep 08 &  2040 &     0 &	- & 2,5\\
Mrk 6	&	0.01951$^\mathrm{a}$	&	10.59	&  10.09 & 	2.14	&	593
&	1.08	&	2006 Nov 21	&	6000	&	0
&	I & 1,4,7\\
Mrk 10	&	0.02925	&	10.78	&  10.38 & 	4.14	&	620
&	2.18	&	2006 Nov 21	&	3600	&	0
&	- & 1	\\
Mrk 79	&	0.02221$^\mathrm{a}$	&	10.85	&  10.33 & 	3.68	&	356
&	0.39	&	2006 Nov 22	&	4800	&	0
&	S & 1,4,8\\
Mrk 110	&	0.03513$^\mathrm{b}$	&	\nodata	&	\nodata	& \nodata &	147
&	0.00	&	2006 Nov 22	&	3600	&	0
&	- & -	\\
Mrk 352	&	0.01486$^\mathrm{a}$	&	\nodata	&	\nodata	& \nodata &	300
&	0.60	&	2006 Nov 22	&	7200	&	0
&	- & 2	\\
Mrk 359	&	0.01694	&	10.35	&  9.96 &	1.59	&	235
&	0.55	&	2005 Sep 05	&	2700	&	0
&	S & 2,5,8\\
Mrk 382	&	0.03348$^\mathrm{b}$	&	10.72	&  10.00 &	1.73	&
\nodata	&	0.00	&	2006 Nov 23	&	4800	&
0	&	B,S & 3,7,8\\
Mrk 477	&	0.03744$^\mathrm{b}$	&	11.14	&  10.72 & 	9.01	&
\nodata	&	0.00	&	2006 Mar 31	&	3926	&
90	&	- & 4	\\
Mrk 506	&	0.04303$^\mathrm{a}$	&	10.69	&  10.12 &	2.30	&	170
&	2.96	&	2005 Sep 08	&	4830	&	27
&	- & 1	\\
Mrk 595	&	0.02739$^\mathrm{b}$	&	10.64	&  10.28 &	3.31	&	444
&	1.41	&	2006 Nov 23	&	6000	&	0
&	S & 3,8\\
Mrk 1018	&	0.04263	&	\nodata	&	\nodata	&
\nodata & 414	&	1.58	&	2006 Nov 23	&	6000	&
0	&	- & 1	\\
Mrk 1126	&	0.01057	&	\nodata	&	\nodata	&
\nodata & 540	&	1.70	&	2006 Nov 22	&	7200	&
0	&	S & 1,8	\\
NGC 788	&	0.01350	&	10.04	&  9.38 & 	0.41	&	226
&	3.20	&	2005 Sep 04	&	2700	&	0
&	S & 1,6	\\
NGC 1019 &	0.02460$^\mathrm{a}$	&	10.43	&  9.97 &	1.62	&
178	&	2.63	&	2006 Nov 22	&	6000	&
0	&	B,R,S & 2,7,8	\\
NGC 7603 &	0.02956	&	10.78	&  10.41 &	4.43	&
313	&	0.45	&	2006 Nov 23	&	7200	&
0	&	R,S & 1,4,8	\\
UGC 3223 &	0.01567	&	10.37	&  10.09 &	2.13	&
263	&	1.66	&	2006 Nov 22	&	6000	&
0	&	S & 1,4,8	\\
\enddata
\tablecomments{Col.(2): Heliocentric redshift.  All redshifts are
  based on stellar measurements except (a) HI 21-cm measurements, or
  (b) measured from the emission lines in our data.  Col.(3): Infrared
  luminosity, in logarithmic units of \lsun.  Col.(4): Far-infrared
  luminosity, in logarithmic units of \lsun (see Section \ref{sfr}).
  Col.(5): Star formation rate, computed from the far-infrared
  luminosity (Section \ref{sfr}).  Col.(6): Circular velocity, equal
  to $\sqrt{2}\sigma$, $v_{rot}$, or the quadratic combination if both
  are available.  Col.(7): Rest-frame equivalent width of \nad\, as
  computed from our model fits.  Col.(8): Observing dates (Section
  \ref{obs}).  Instrument used was the R-C Spectrograph on the KPNO
  4m.  Col.(9): Total exposure time in seconds.  Col.(10): Slit
  position angle.  Col.(11): Letters indicate dust structure around
  nucleus (see Section \ref{nuc}): (B) nuclear dust bar, (F) dust
  filaments, (I) irregular dust, (R) nuclear dust ring, (S) nuclear
  dust spiral.  Col.(12): Reference.}  

\tablerefs{(1) \citealt{nw95}; (2) \citealt{sob05}; (3)
  \citealt{w92a}; (4) IRAS Faint Source Catalog; (5) IRAS Point Source
  Catalog; (6) \citealt{m03a}; (7) \citealt{mgt98}; (8)
  \citealt{dck06}.}
\end{deluxetable}

\begin{deluxetable}{lcrrccrr}
%\rotate
\tabletypesize{\footnotesize}
\tablecaption{Properties of individual velocity components \label{compprop}}
\tablewidth{0pt}
\tablehead{
\colhead{} & \colhead{$\lambda_{1,c}$} & \colhead{$\Delta v$} & \colhead{$b$} & \colhead{$\tau_{1,c}$} & \colhead{\cf} & \colhead{$N$(\ion{Na}{1})} & \colhead{$N$(H)}\\
\colhead{Name} & \colhead{(\AA)} & \colhead{(km s$^{-1}$)} & \colhead{(\kms)} & \colhead{} & \colhead{} & \colhead{(cm $^{-2}$)} & \colhead{(cm $^{-2}$)}\\
\colhead{(1)} & \colhead{(2)} & \colhead{(3)} & \colhead{(4)} & \colhead{(5)} & \colhead{(6)} & \colhead{(7)} & \colhead{(8)}
}
\startdata
\multicolumn{8}{c}{\bf{Seyfert 2s}}\\
\tableline
Akn 79	& 	5999.47	& 	-44 $\pm$ 5	& 	85 $\pm$ 19
& 	1.07 $^{+0.50}_{-0.21}$	& 	0.12 $^{+0.06}_{-0.02}$	&
13.44 $^{+0.28}_{-0.20}$	& 	21.08 $^{+0.43}_{-0.43}$	\\
\nodata	& 	5999.86	& 	-25 $\pm$ 7	& 	280 $\pm$ 22
& 	0.27 $^{+0.04 }_{-0.04}$	& 	0.48 $^{+ 0.07
}_{-0.06}$	& 	13.42  $^{+0.12}_{-0.12}$	& 	21.06 $^{+0.18}_{-0.18}$	\\
Arp 107A	& 	6102.90	& 	54 $\pm$ 24	& 	323
$\pm$ 59	& 	0.07 $^{+0.05 }_{-0.01}$	& 	0.92
$^{+0.04 }_{-0.01}$	& 	12.94  $^{+0.25}_{-0.09}$	&
20.58 $^{+0.25}_{-0.09}$	\\
Mrk 348	& 	5988.26	& 	65 $\pm$ 4	& 	261 $\pm$ 25
& 	0.06 $^{+0.01 }_{-0.01}$	& 	1.00 $^{+0.01
}_{-0.01}$	& 	12.74  $^{+0.08}_{-0.06}$	& 	20.38 $^{+0.08}_{-0.06}$	\\
Mrk 622	& 	6033.19	& 	-138 $\pm$ 10	& 	416 $\pm$ 62
& 	0.09 $^{+0.04 }_{-0.01}$	& 	0.55 $^{+0.23
}_{-0.02}$	& 	13.13 $^{+0.16}_{-0.07}$	& 	20.77  $^{+0.16}_{-0.07}$	\\
Mrk 686	& 	5981.63	& 	16 $\pm$ 5	& 	287 $\pm$ 34	& 	0.10 $^{+0.04 }_{-0.01}$	& 	1.00 $^{+0.01 }_{-0.01}$	& 	13.05 $^{+0.25}_{-0.11}$	& 	20.69 $^{+0.31}_{-0.17}$	\\
Mrk 1157	& 	5986.67	& 	3 $\pm$ 2	& 	211
$\pm$ 20	& 	0.83 $^{+0.17 }_{-0.13}$	& 	0.19
$^{+0.04 }_{-0.03}$	& 	13.79 $^{+0.17}_{-0.13}$	&
21.43 $^{+0.23}_{-0.20}$	\\
NGC 1358	& 	5977.92	& 	35 $\pm$ 6	& 	320 $\pm$ 19	& 	0.43 $^{+0.05 }_{-0.05}$	& 	0.37 $^{+0.04 }_{-0.04}$	& 	13.69 $^{+0.08}_{-0.08}$	& 	21.33 $^{+0.15}_{-0.15}$	\\
NGC 1667	& 	5986.79	& 	-12 $\pm$ 84	& 	343 $\pm$ 19	& 	0.32 $^{+0.03 }_{-0.03}$	& 	0.46 $^{+0.05 }_{-0.05}$	& 	13.59 $^{+0.06}_{-0.06}$	& 	21.23 $^{+0.13}_{-0.13}$	\\
NGC 3362	& 	6062.43	& 	84 $\pm$ 38	& 	323 $\pm$ 92	& 	0.05 $^{+0.07 }_{-0.01}$	& 	1.00 $^{+0.01 }_{-0.01}$	& 	12.83 $^{+0.39}_{-0.16}$	& 	20.47 $^{+0.39}_{-0.16}$	\\
NGC 3786	& 	5947.88	& 	-148 $\pm$ 5	& 	34
$\pm$ 3	& 	5.00 $^{+0.36 }_{-2.00}$	& 	0.10 $^{+0.01
}_{-0.04}$	& 	$>$13.76  $^{+0.05}_{-0.23}$	& 	$>$21.40 $^{+0.05}_{-0.23}$	\\
\nodata	& 	5951.19	& 	19 $\pm$ 2	& 	239 $\pm$ 14
& 	0.34 $^{+0.04 }_{-0.03}$	& 	0.44 $^{+0.05
}_{-0.04}$	& 	$>$13.45  $^{+0.03}_{-0.21}$	&
$>$21.09  $^{+0.03}_{-0.21}$	\\
NGC 4388	& 	5948.05	& 	51 $\pm$ 17	& 	125
$\pm$ 5	& 	0.98 $^{+0.07 }_{-0.07}$	& 	0.32 $^{+ 0.02
}_{- 0.02}$	& 	13.60  $^{+0.04}_{-0.04}$ & 	21.24 $^{+0.04}_{-0.04}$\\
NGC 5252	& 	6034.47	& 	39 $\pm$ 10	& 	375
$\pm$ 23	& 	0.10 $^{+0.02 }_{-0.01}$	& 	1.00
$^{+0.01 }_{-0.01}$	& 	13.14  $^{+0.15}_{-0.07}$	& 	20.78 $^{+0.21}_{-0.13}$	\\
NGC 5728	& 	5952.30	& 	-11 $\pm$ 1	& 	262
$\pm$ 26	& 	0.27 $^{+0.06 }_{-0.03}$	& 	0.37
$^{+0.08 }_{-0.04}$	& 	13.40 $^{+0.17}_{-0.13}$	& 	21.04 $^{+0.23}_{-0.19}$	\\
NGC 7319	& 	6026.74	& 	-133 $\pm$ 19	& 	100 $\pm$ 1	& 	3.00 $^{+0.24 }_{-2.12}$	& 	0.20 $^{+0.02 }_{-0.14}$	& 	13.34 $^{+0.04}_{-0.10}$	& 	20.98 $^{+0.04}_{-0.10}$	\\
\nodata	& 	6026.82	& 	-130 $\pm$ 1	& 	450 $\pm$ 1
& 	0.46 $^{+0.06 }_{-0.06}$	& 	0.49 $^{+0.06
}_{-0.06}$	& 	13.81 $^{+0.05}_{-0.12}$	& 	21.45 $^{+0.05}_{-0.12}$	\\
NGC 7672	& 	5976.55	& 	-25 $\pm$ 24	& 	98 $\pm$ 6	& 	1.12 $^{+0.17 }_{-0.07}$	& 	0.24 $^{+0.04 }_{-0.01}$	& 	13.53 $^{+0.11}_{-0.06}$	& 	21.17 $^{+0.18}_{-0.03}$	\\
NGC 7682	& 	6000.21	& 	99 $\pm$ 3	& 	228 $\pm$ 21	& 	0.49 $^{+0.09 }_{-0.08}$	& 	0.25 $^{+0.05 }_{-0.04}$	& 	13.59 $^{+0.08}_{-0.09}$	& 	21.23 $^{+0.08}_{-0.09}$	\\
UGC 3995	& 	5992.99	& 	127 $\pm$ 2	& 	268
$\pm$ 15	& 	0.13 $^{+0.02 }_{-0.01}$	& 	1.00
$^{+0.01 }_{-0.01}$	& 	13.09 $^{+0.07}_{-0.03}$	& 	20.73 $^{+0.07}_{-0.03}$	\\
\tableline 
\multicolumn{8}{c}{\bf{Seyfert 1s}}\\ 
\tableline
Akn 202	& 	6069.22	& 	113 $\pm$ 11	& 	129 $\pm$ 32
& 	1.02 $^{+0.23 }_{-0.05}$	& 	0.05 $^{+0.01
}_{-0.01}$	& 	13.67 $^{+0.12}_{-0.13}$	& 	21.31 $^{+0.12}_{-0.13}$	\\
Mrk 6	& 	6008.03	& 	-229 $\pm$ 1	& 	25 $\pm$ 1
& 	1.46 $^{+0.03 }_{-0.03}$	& 	0.48 $^{+0.01
}_{-0.01}$	& 	13.12 $^{+0.17}_{-0.29}$	& 	20.76 $^{+0.17}_{-0.29}$	\\
\nodata	& 	5992.12	& 	-1024 $\pm$ 1	& 	16 $\pm$ 11	& 	4.75 $^{+0.23 }_{-0.89}$	& 	0.06 $^{+0.01 }_{-0.01}$	& 	13.44 $^{+0.20}_{-0.31}$	& 	21.08 $^{+0.20}_{-0.31}$	\\
Mrk 10	& 	6070.59	& 	26 $\pm$ 7	& 	305 $\pm$ 34	& 	0.17 $^{+0.06 }_{-0.02}$	& 	0.45 $^{+0.15 }_{-0.04}$	& 	13.28 $^{+0.23}_{-0.11}$	& 	20.92 $^{+0.30}_{-0.17}$	\\
Mrk 79	& 	6028.25	& 	-15 $\pm$ 7	& 	112 $\pm$ 23	& 	0.92 $^{+0.32 }_{-0.31}$	& 	0.06 $^{+0.06 }_{-0.02}$	& 	13.57 $^{+0.25}_{-0.24}$	& 	21.21 $^{+0.31}_{-0.30}$	\\
Mrk 352	& 	5986.09	& 	45 $\pm$ 8	& 	111 $\pm$ 25	& 	0.16 $^{+0.04 }_{-0.01}$	& 	0.16 $^{+0.03 }_{-0.01}$	& 	12.82 $^{+0.07}_{-0.42}$	& 	20.46 $^{+0.07}_{-0.42}$	\\
Mrk 359	& 	5997.12	& 	-17 $\pm$ 11	& 	147 $\pm$ 28	& 	0.21 $^{+0.12 }_{-0.03}$	& 	0.20 $^{+0.12 }_{-0.03}$	& 	13.03 $^{+0.30}_{-0.17}$& 	20.67 $^{+0.36}_{-0.23}$	\\
Mrk 506	& 	6154.09	& 	135 $\pm$ 7	& 	296 $\pm$ 32	& 	0.17 $^{+0.05 }_{-0.01}$	& 	0.64 $^{+0.18 }_{-0.05}$	& 	13.26 $^{+0.12}_{-0.06}$	& 	20.90 $^{+0.12}_{-0.06}$	\\
Mrk 595	& 	6059.26	& 	8 $\pm$ 5	& 	230 $\pm$ 21	& 	0.15 $^{+0.03 }_{-0.01}$	& 	0.44 $^{+0.08 }_{-0.03}$	& 	13.09 $^{+0.15}_{-0.05}$	& 	20.73 $^{+0.22}_{-0.12}$	\\
Mrk 1018	& 	6150.34	& 	67 $\pm$ 5	& 	379 $\pm$ 31	& 	0.12 $^{+0.03 }_{-0.01}$	& 	0.36 $^{+0.09 }_{-0.03}$	& 	13.22 $^{+0.10}_{-0.05}$	& 	20.86 $^{+0.10}_{-0.05}$	\\
Mrk 1126	& 	5962.64	& 	138 $\pm$ 3	& 	204 $\pm$ 11	& 	0.08 $^{+0.01 }_{-0.01}$	& 	1.00 $^{+0.01 }_{-0.01}$	& 	12.79 $^{+0.06}_{-0.03}$	& 	20.43 $^{+0.06}_{-0.03}$	\\
NGC 788	& 	5977.35	& 	0 $\pm$ 3	& 	253 $\pm$ 61	& 	0.72 $^{+0.06 }_{-0.08}$	& 	0.28 $^{+0.02 }_{-0.03}$	& 	13.81 $^{+0.19}_{-0.19}$	& 	21.45 $^{+0.25}_{-0.25}$	\\
NGC 1019	& 	6044.02	& 	68 $\pm$ 4	& 	253 $\pm$ 23	& 	0.31 $^{+0.05 }_{-0.04}$	& 	0.41 $^{+0.07 }_{-0.05}$	& 	13.44 $^{+0.07}_{-0.07}$	& 	21.08 $^{+0.07}_{-0.07}$	\\
NGC 7603	& 	6073.78	& 	93 $\pm$ 11	& 	497 $\pm$ 45	& 	0.06 $^{+0.04 }_{-0.05}$	& 	0.14 $^{+0.10 }_{-0.11}$	& 	13.06 $^{+0.22}_{-0.70}$	& 	20.70 $^{+0.22}_{-0.70}$	\\
UGC 3223	& 	5989.19	& 	-39 $\pm$ 1	& 	81 $\pm$ 10	& 	5.00 $^{+0.77 }_{-1.89}$	& 	0.08 $^{+0.01 }_{-0.03}$	& 	14.16 $^{+0.17}_{-0.17}$	& 	21.80 $^{+0.17}_{-0.17}$	\\
\nodata	& 	5991.73	& 	88 $\pm$ 1	& 	98 $\pm$ 22
& 	0.92 $^{+0.39 }_{-0.20}$	& 	0.16 $^{+0.07 }_{-0.04}$	&
13.50 $^{+0.12}_{-0.13}$	& 	21.14 $^{+0.12}_{-0.13}$	\\
\enddata
\tablecomments{Col.(2): Redshifted heliocentric wavelength, in vacuum,
  of the \ion{Na}{1}~D$_1$ $\lambda5896$ line.  Col.(3): Velocity
  relative to systemic.  Negative velocities are blueshifted, positive
  are redshifted.  Components with $\Delta v < -50$ \kms~and $|\Delta
  v|$ $>$ 2$\delta$($\Delta v$) are assumed to be outflowing; those
  with $\Delta v > 50$ \kms~and $|\Delta v|$ $>$ 2$\delta$($\Delta v$)
  are assumed to be inflowing .  Col.(4): Doppler parameter.  Col.
  (5): Central optical depth of the \ion{Na}{1}~D$_1$ $\lambda5896$
  line; the optical depth of the D$_2$ line is twice this value.
  Col.(6): Covering fraction of the gas.  Col.(7-8): Logarithm of
  column density of \ion{Na}{1}~and H, respectively.}
\end{deluxetable}

\begin{deluxetable}{lrrcccccc}
%\rotate
\tabletypesize{\footnotesize}
\tablecaption{Outflow: Individual objects \label{outindiv}}
\tablewidth{0pt}
\tablehead{
\colhead{} & \colhead{\dvmax} & \colhead{$M$} & \colhead{$dM/dt$} & \colhead{$p$} & \colhead{$dp/dt$} & \colhead{$E$} & \colhead{$dE/dt$}\\
\colhead{Name} & \colhead{(\kms)} & \colhead{($M_\odot$)} &
\colhead{(\smpy)} & \colhead{(dyn s)} & \colhead{(dyn)} & \colhead{(ergs)}
& \colhead{(ergs s$^{-1}$)}\\
\colhead{(1)} & \colhead{(2)} & \colhead{(3)} & \colhead{(4)} & \colhead{(5)} & 
\colhead{(6)} & \colhead{(7)} & \colhead{(8)} 
}
\startdata
\multicolumn{8}{c}{\bf{Seyfert 2s}}\\
\tableline
Mrk 622	&	-484	&	8.76	&	1.22	&	49.2	&	34.2	&	56.1	&	41.0	\\
NGC 3786	&	-176	&	8.64	&	1.12	&	49.1	&	34.1	&	56.0	&	41.0	\\
NGC 7319	&	-217	&	8.52	&	0.95	&	48.9	&	33.9	&	55.8	&	40.7	\\
\nodata	&	-504	&	9.39	&	1.85	&	49.8	&	34.8	&	56.7	&	41.7	\\
\tableline
\multicolumn{8}{c}{\bf{Seyfert 1s}}\\
\tableline
Mrk 6 	&	-250	&	8.69	&	1.36	&	49.4	&	34.5	&	56.4	&	41.6	\\
\nodata	&	-1037	&	8.08	&	1.40	&	49.4	&	35.2	&	57.1	&	42.9	\\
\enddata
\tablecomments{Col.(2): Maximum velocity in the outflow, $\dvmax
  \equiv \Delta v - \mathrm{FWHM}/2$.  Col .(3): Log of total
  outflowing mass.  Col.(4): Log of mass outflow rate.  Col.  (5): Log
  of total momentum of outflow.  Col.(6): Log of momentum outflow
  rate.  Col.(7): Log of total kinetic energy of outflow.  Col.(8):
  Log of kinetic energy outflow rate.  Note that all values are
  calculated using $\Omega$ $\sim$ 0.1.}
\end{deluxetable}

\begin{deluxetable}{lcccc}
\tabletypesize{\footnotesize}
\tablecaption{Outflow: Average properties \label{avgpropout}}
\tablewidth{0pt}
\tablehead{
\colhead{Quantity} & \colhead{IR-Faint Seyfert 2s}  & \colhead{IR-Faint
  Seyfert 1s} & \colhead{IR-Lum. Seyfert~2s} & \colhead{IR-Lum. Seyfert~1s} \\
\colhead{(1)} & \colhead{(2)} & \colhead{(3)} & \colhead{(4)} & \colhead{(5)}
}
\startdata
Number of galaxies	&	17	&	18	&	20	&	6\\
Detection rate (\%)	&	18 $\pm$ 9	&	6 $\pm$ 6	&	45 $\pm$ 11	&	50 $\pm$ 20\\
\tableline
\multicolumn{5}{c}{\bf{Galaxy Properties}}\\
\tableline
$z$	&	0.018 $\pm$ 0.01	&	0.025 $\pm$ 0.01	&	0.148 $\pm$ 0.11	&	0.150 $\pm$ 0.09 	\\
log(\lfir/\lsun)	&	10.02 $\pm$ 0.35	&	10.14 $\pm$ 0.23	&	11.86 $\pm$ 0.31	&	12.13 $\pm$ 0.1	\\
SFR (\smpy)	&	2.39 $\pm$ 1.92	&	2.99 $\pm$ 2.13	&	118 $\pm$ 109	&	164 $\pm$ 75	\\
$\Delta v$ (\kms) &     -137 $\pm$ 8 &  -627 $\pm$ 562 & -322 $\pm$
388 & -4942 $\pm$ 2831\\
$\Delta v_{max}$ (\kms)	&	-345 $\pm$ 173	&	-643 $\pm$ 556	&	-618 $\pm$ 422	&	-5210 $\pm$ 4306	\\
log[$N$(\nad)/cm s$^{-2}$]	&	13.51 $\pm$ 0.33	&	13.28 $\pm$ 0.23	&	13.5 $\pm$ 0.7	&	14.5 $\pm$ 0.8	\\
log[$N$(H)/cm s$^{-2}$]	& 21.15 $\pm$ 0.33	&	20.92 $\pm$ 0.23	&	20.9 $\pm$ 0.7	&	21.8 $\pm$ 0.8	\\
\tableline
\multicolumn{5}{c}{\bf{Velocity Component Properties}}\\
\tableline
$\tau$	&	0.76 $^{+0.1}_{-0.7}$	&	1.01 $^{+0.5}_{-0.1}$	&	0.27 $^{+0.7}_{-0.2}$	&	0.69 $^{+1.6}_{-0.5}$	\\
$b$ (\kms)	&	250 $\pm$ 214	&	21 $\pm$ 6	&	232 $\pm$ 182	&	87 $\pm$ 130	\\
\cf	&	0.34 $^{+0.1}_{-0.1}$	&	0.27 $^{+0.1}_{-0.1}$	&	0.42 $^{+0.5}_{-0.2}$	&	0.67 $^{+1.2}_{-0.4}$	\\
\enddata
\tablecomments{``IR-Lum.'' refers to the IR-luminous data of RVS05c.
  For most quantities we list the mean and 1$\sigma$ dispersions,
  under the assumption of a Gaussian distribution in the log of the
  quantity.  Statistics for all quantities except $z$,~\lfir, and SFR
  are computed only for outflowing velocity components.  Note that the
  entries under IR-Faint Seyfert 2s (1s) corresponds to only three
  (one) objects.}
\end{deluxetable}

\begin{deluxetable}{lccrccccc}
%\rotate
\tabletypesize{\footnotesize}
\tablecaption{Inflow: Individual objects \label{inindiv}}
\tablewidth{0pt}
\tablehead{
\colhead{} & \colhead{\dvmax} & \colhead{$M$} & \colhead{$dM/dt$} & \colhead{$p$} & \colhead{$dp/dt$} & \colhead{$E$} & \colhead{$dE/dt$}\\
\colhead{Name} & \colhead{(\kms)} & \colhead{($M_\odot$)} &
\colhead{(\smpy)} & \colhead{(dyn s)} & \colhead{(dyn)} & \colhead{(ergs)}
& \colhead{(ergs s$^{-1}$)}\\
\colhead{(1)} & \colhead{(2)} & \colhead{(3)} & \colhead{(4)} & \colhead{(5)} & 
\colhead{(6)} & \colhead{(7)} & \colhead{(8)} 
}
\startdata
\multicolumn{8}{c}{\bf{Seyfert 2s}}\\
\tableline
Arp 107A	&	323	&	7.39	&	0.15	&	48.4	&	32.7	&	53.9	&	39.2	\\
Mrk 348	        &	282	&	7.22	&	0.04	&	47.3	&	32.7	&	53.8	&	39.2	\\
NGC 3362	&	352	&	7.32	&	0.27	&	47.6	&	33.0	&	54.2	&	39.6	\\
NGC 4388	&	155     &	7.59	&	0.31	&	47.6	&	32.8	&	54.0	&	39.2	\\
NGC 7682	&	289	&	7.48	&	0.49	&	47.8	&	33.3	&	54.5	&	40.0	\\
UGC 3995	&	350	&	7.58	&	0.69	&	48.0	&	33.6	&	54.8	&	40.4	\\
\tableline
\multicolumn{8}{c}{\bf{Seyfert 1s}}\\
\tableline
Akn 202	        &	220	&	6.89	&	-0.04	&	47.3	&	32.8	&	54.0	&	39.6	\\
Mrk 506	        &	381	&	7.55	&	0.69	&	48.0	&	33.6	&	54.8	&	40.5	\\
Mrk 1018	&	382	&	7.27	&	0.10	&	47.4	&	32.7	&	53.9	&	39.3	\\
Mrk 1126	&	308	&	7.27	&	0.43	&	47.7	&	33.4	&	54.6	&	40.2	\\
NGC 1019	&	279	&	7.54	&	0.39	&	47.7	&	33.0	&	54.2	&	39.6	\\
NGC 7603	&	507	&	6.70	&	-0.32	&	47.0	&	32.5	&	53.6	&	39.1	\\
UGC 3223	&	169	&	7.19	&	0.15	&	47.4	&	32.9	&	54.1	&	39.5	\\
\enddata

\tablecomments{All inflow dynamics calculations assume an absorber
  radius of 1 kpc.  Col.(2): Maximum velocity of the inflow, $\dvmax
  \equiv \Delta v + \mathrm{FWHM}/2$.  Col.(3): Log of total inflowing
  mass.  Col.(4): Log of mass inflow rate.  Col.  (5): Log of total
  momentum of inflow.  Col.(6): Log of momentum inflow rate.  Col.(7):
  Log of total kinetic energy of inflow.  Col.(8): Log of kinetic
  energy inflow rate.}
\end{deluxetable}

\begin{deluxetable}{lcccc}
\tabletypesize{\footnotesize}
\tablecaption{Inflow: Average properties \label{avgpropin}}
\tablewidth{0pt}
\tablehead{
\colhead{Quantity} & \colhead{IR-Faint Seyfert 2s}  & \colhead{IR-Faint Seyfert 1s} \\
\colhead{(1)} & \colhead{(2)} & \colhead{(3)}}
\startdata
Number of galaxies	&	17	&	18	\\
Detection rate (\%)	&	35 $\pm$ 11	&	39 $\pm$ 12	\\
\tableline
\multicolumn{3}{c}{\bf{Galaxy Properties}}\\
\tableline
$z$	&	0.018 $\pm$ 0.01	&	0.025 $\pm$ 0.01	\\
log(\lfir/\lsun)	&	10.02 $\pm$ 0.35	&	10.14 $\pm$ 0.23	\\
SFR (\smpy)	&	2.39 $\pm$ 1.92	&	2.99 $\pm$ 2.13	\\
$\Delta v$ (\kms) & 80 $\pm$ 29 & 100 $\pm$ 29\\
$\Delta v_{max}$ (\kms)	&	291 $\pm$ 37	&	321 $\pm$ 29	\\
log[$N$(\nad)/cm s$^{-2}$]	&	13.13 $\pm$ 0.38	&	13.28 $\pm$ 0.29	\\
log[$N$(H)/cm s$^{-2}$]	&	20.77 $\pm$ 0.38	&	20.92 $\pm$ 0.29	\\
\tableline
\multicolumn{3}{c}{\bf{Velocity Component Properties}}\\
\tableline
$\tau$	&	0.3 $^{+0.13}_{-0.06}$	&	0.38 $^{+0.05}_{-0.03}$	\\
$b$ (\kms)	&	254 $\pm$ 74	&	265 $\pm$ 140	\\
\cf	&	0.75 $^{+0.09}_{-0.05}$	&	0.39 $^{+0.02}_{-0.02}$	\\
\enddata
\tablecomments{For most quantities we list the mean and 1$\sigma$
  dispersions, under the assumption of a Gaussian distribution in the
  log of the quantity.  Statistics for all quantities except $z$,~\lfir,
  and SFR are computed only for inflowing velocity components.}
\end{deluxetable}

\begin{deluxetable}{lrr}
\tabletypesize{\footnotesize}
\tablecaption{Statistical comparisons of kinematic parameters: IR-faint and IR-luminous Seyfert galaxies \label{stattests}}
\tablewidth{0pt}
\tablehead{
\colhead{Samples} & \colhead{$P$(null, K-S)}  & \colhead{$P$(null, Kuiper)}} 
\startdata
\multicolumn{3}{c}{\boldmath{$\Delta v$}}\\
\tableline
IR-Faint Seyfert 1s vs IR-Faint Seyfert 2s	&	0.76	&	0.69	\\
IR-Faint Seyfert 1s vs IR-Luminous Seyfert 1s	&	\textbf{$<$0.01}	&	\textbf{$<$0.01}	\\
IR-Faint Seyfert 2s vs IR-Luminous Seyfert 2s	&	\textbf{0.04}	&	\textbf{0.07}	\\
IR-Faint Seyferts vs IR-Luminous Seyferts       &
\textbf{$<$0.01} & \textbf{$<$0.01} \\
\tableline
\multicolumn{3}{c}{\boldmath{\dvmax}}\\
\tableline
IR-Faint Seyfert 1s vs IR-Faint Seyfert 2s	&	0.84	&	0.95	\\
IR-Faint Seyfert 1s vs IR-Luminous Seyfert 1s	&	\textbf{$<$0.01}	&	\textbf{$<$0.01}	\\
IR-Faint Seyfert 2s vs IR-Luminous Seyfert 2s	&
\textbf{$<$0.01}	&	\textbf{0.03}	\\
IR-Faint Seyferts vs IR-Luminous Seyferts       &
\textbf{$<$0.01} & \textbf{$<$0.01} \\
\tableline
\multicolumn{3}{c}{\bf{Doppler Parameter}}\\
\tableline
IR-Faint Seyfert 1s vs IR-Faint Seyfert 2s	&	0.33	&	0.71	\\
IR-Faint Seyfert 1s vs IR-Luminous Seyfert 1s	&	\textbf{0.09}	&	\textbf{0.06}	\\
IR-Faint Seyfert 2s vs IR-Luminous Seyfert 2s	&	0.28	&
0.72	\\
IR-Faint Seyferts vs IR-Luminous Seyferts       &
0.23 & 0.41 \\
\enddata
\tablecomments{$P$(null) is the probability that the two listed
  distributions are taken from the same parent population.
  Categories which have $P$(null) $<$ 0.1 for both tests are
  printed in bold.  Values are based on all absorption features.}
\end{deluxetable}

\begin{deluxetable}{lrr}
\tabletypesize{\footnotesize}
\tablecaption{Statistical comparisons of kinematic
  parameters: All galaxies \label{stattests2a}}
\tablewidth{0pt}
\tablehead{
\colhead{Samples} & \colhead{$P$(null, K-S)}  & \colhead{$P$(null, Kuiper)}} 
\startdata
\multicolumn{3}{c}{\boldmath{$\Delta v$}}\\
\tableline
Seyfert 1s vs Starburst ULIRGs \& LIRGs & \textbf{$<$0.01} & \textbf{$<$0.01}\\
Seyfert 2s vs Starburst ULIRGs \& LIRGs & \textbf{0.03}  & \textbf{0.09} \\
Seyfert 1s vs Seyfert 2s                & \textbf{$<$0.01} & \textbf{$<$0.01}\\
\tableline
\multicolumn{3}{c}{\boldmath{\dvmax}}\\
\tableline
Seyfert 1s vs Starburst ULIRGs \& LIRGs & \textbf{$<$0.01} & \textbf{$<$0.01}\\
Seyfert 2s vs Starburst ULIRGs \& LIRGs & \textbf{0.02}  & \textbf{0.02}\\
Seyfert 1s vs Seyfert 2s                & \textbf{$<$0.01} & \textbf{$<$0.01}\\
\tableline
\multicolumn{3}{c}{\bf{Doppler Parameter}}\\
\tableline
Seyfert 1s vs Starburst ULIRGs \& LIRGs & 0.21 & 0.03\\
Seyfert 2s vs Starburst ULIRGs \& LIRGs & \textbf{$<$0.01} & \textbf{$<$0.01}\\
Seyfert 1s vs Seyfert 2s                & \textbf{0.02}  & \textbf{$<$0.01}\\
\enddata
\tablecomments{$P$(null) is the probability that the two listed
  distributions are taken from the same intrinsic distribution.
  Categories which have $P$(null) $<$ 0.1 for both tests are printed
  in bold.  Values are based on all absorption features, both
  inflowing and outflowing.  Note that Seyfert 1 and Seyfert 2
  galaxies include both IR-Faint Seyferts from the present study, as well
  as IR-Luminous Seyferts from RVS05c.  Starburst data are taken from RVS05b.}
\end{deluxetable}

\begin{deluxetable}{lrr}
\tabletypesize{\footnotesize}
\tablecaption{Statistical comparisons of kinematic parameters: 
Galaxies with outflows \label{stattests2b}}
\tablewidth{0pt}
\tablehead{
\colhead{Samples} & \colhead{$P$(null, K-S)}  & \colhead{$P$(null, Kuiper)}} 
\startdata
\multicolumn{3}{c}{\boldmath{$\Delta v$}}\\
\tableline
Seyfert 1s vs Starburst ULIRGs \& LIRGs & \textbf{$<$0.01} & \textbf{$<$0.01}\\
Seyfert 2s vs Starburst ULIRGs \& LIRGs & 0.96 & 0.84\\
Seyfert 1s vs Seyfert 2s                & \textbf{$<$0.01} & \textbf{$<$0.01}\\
\tableline
\multicolumn{3}{c}{\boldmath{\dvmax}}\\
\tableline
Seyfert 1s vs Starburst ULIRGs \& LIRGs & \textbf{$<$0.01} & \textbf{$<$0.01}\\
Seyfert 2s vs Starburst ULIRGs \& LIRGs & 0.26 & 0.31\\
Seyfert 1s vs Seyfert 2s                & \textbf{$<$0.01} & \textbf{$<$0.01}\\
\tableline
\multicolumn{3}{c}{\bf{Doppler Parameter}}\\
\tableline
Seyfert 1s vs Starburst ULIRGs \& LIRGs & \textbf{$<$0.01} & \textbf{$<$0.01}\\
Seyfert 2s vs Starburst ULIRGs \& LIRGs & 0.24 & 0.28 \\
Seyfert 1s vs Seyfert 2s                & \textbf{0.02} & \textbf{0.01}\\
\enddata
\tablecomments{$P$(null) is the probability that the two listed
  distributions are taken from the same parent population.
  Categories which have $P$(null) $<$ 0.1 for both tests are
  printed in bold.  Values are based on outflowing
  absorption only.  Note that Seyfert 1 and Seyfert 2
  galaxies include both IR-Faint Seyferts from the present study, as well
  as IR-Luminous Seyferts from RVS05c.  Starburst data are taken from RVS05b.}
\end{deluxetable}

\begin{deluxetable}{lcc}
\tabletypesize{\footnotesize}
\tablecaption{Statistical comparisons of kinematic parameters: Seyfert
  1s versus Seyfert 2s with inflows \label{stattests2c}}
\tablewidth{0pt}
\tablehead{
\colhead{Samples} & \colhead{$P$(null, K-S)}  & \colhead{$P$(null, Kuiper)}} 
\startdata
$\Delta v$                & 0.83 & 0.87\\
$\Delta v_{max}$          & 0.72 & 0.89\\
Doppler Parameter         & 0.68 & 0.99\\
\enddata
\tablecomments{$P$(null) is the probability that the two listed
  distributions are taken from the same parent population.  Values are
  based on objects with inflowing absorption only, and include data
  for IR-faint galaxies from this survey and IR-bright galaxies from
  RVS05c.}
\end{deluxetable}

%%%%%%%%%%%
% FIGURES %
%%%%%%%%%%%

\clearpage

\setcounter{figure}{4}

\begin{figure}[t]
\epsscale{0.9}
\plotone{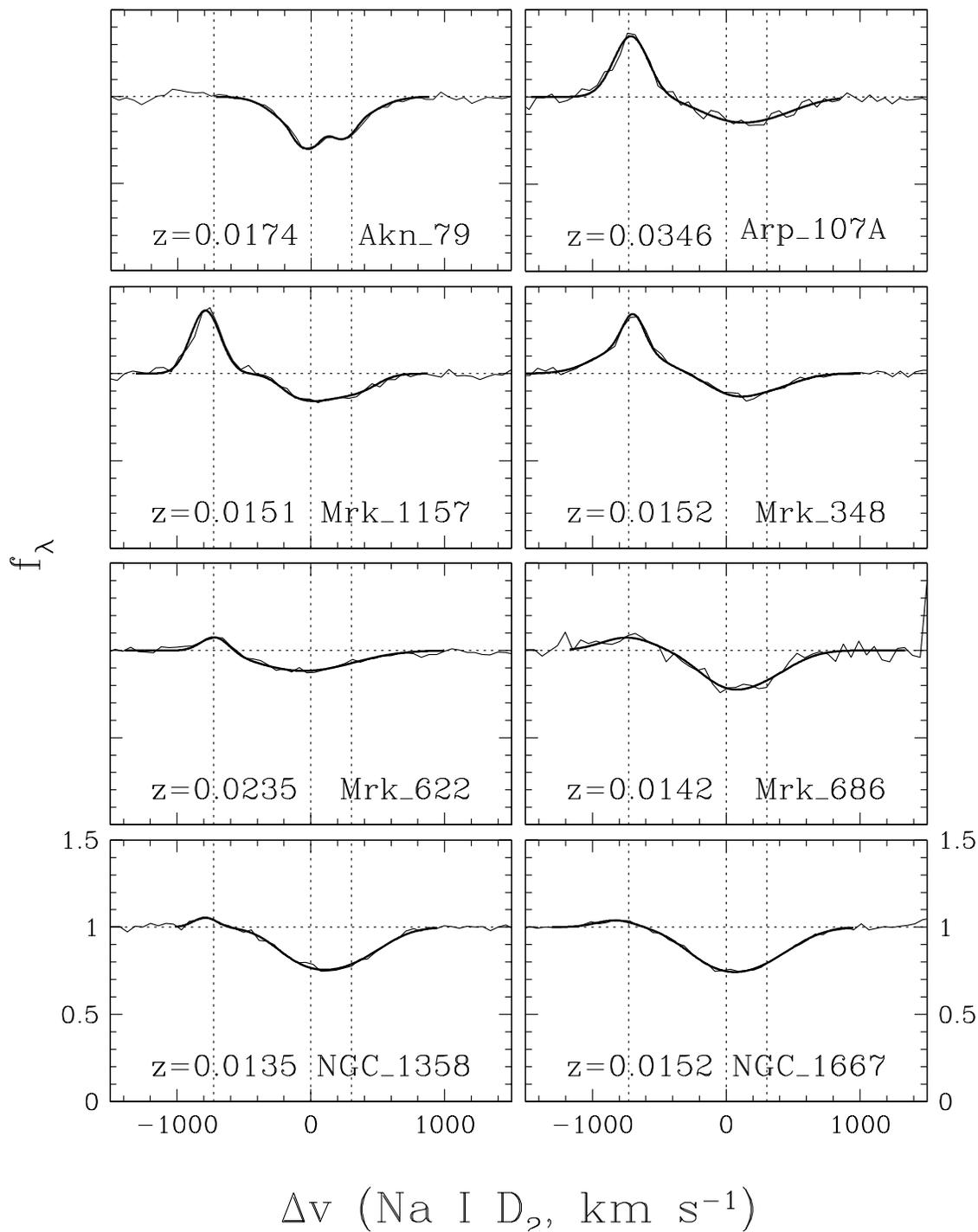}
\caption{Spectra of the \heo emission + \nad absorption region in
  infrared-faint Seyfert 2 galaxies. These spectra are plotted on a
  velocity scale based on the systemic \nad $\lambda$5890 velocity.  The thin lines
  are smoothed original spectra, and thick lines are fits to the data.
  The vertical dotted lines show the locations of the \nad
  $\lambda\lambda$5890, 5896 doublet and \heo $\lambda$5876 emission
  line in the object's rest frame.}
\label{sytspectrai}
\end{figure}

\setcounter{figure}{4}

\begin{figure}[t]
\epsscale{1.0}
\plotone{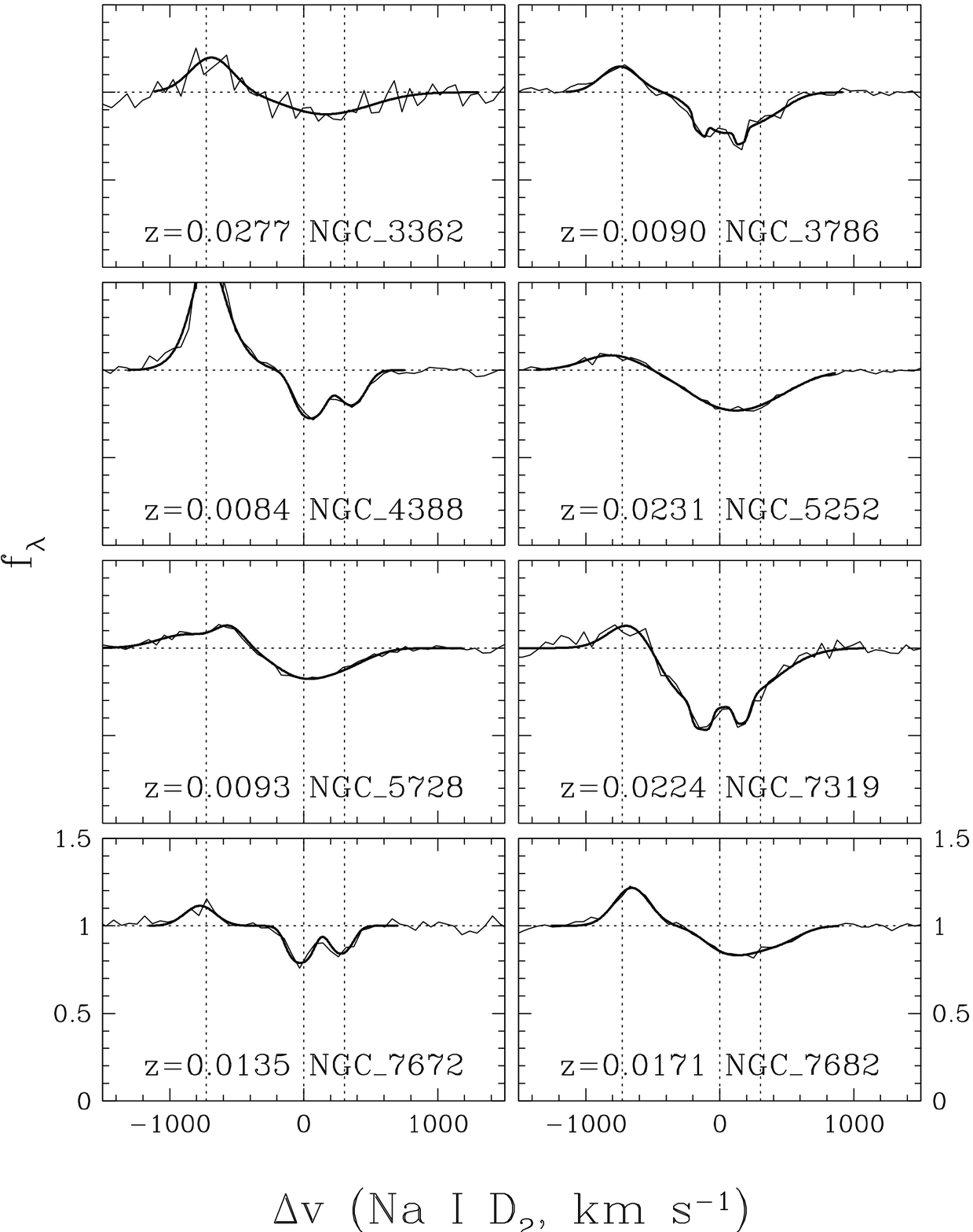}
\caption{$Continued$}
\label{sytspectraii}
\end{figure}

\setcounter{figure}{4}

\begin{figure}[t]
\epsscale{1.0}
\plotone{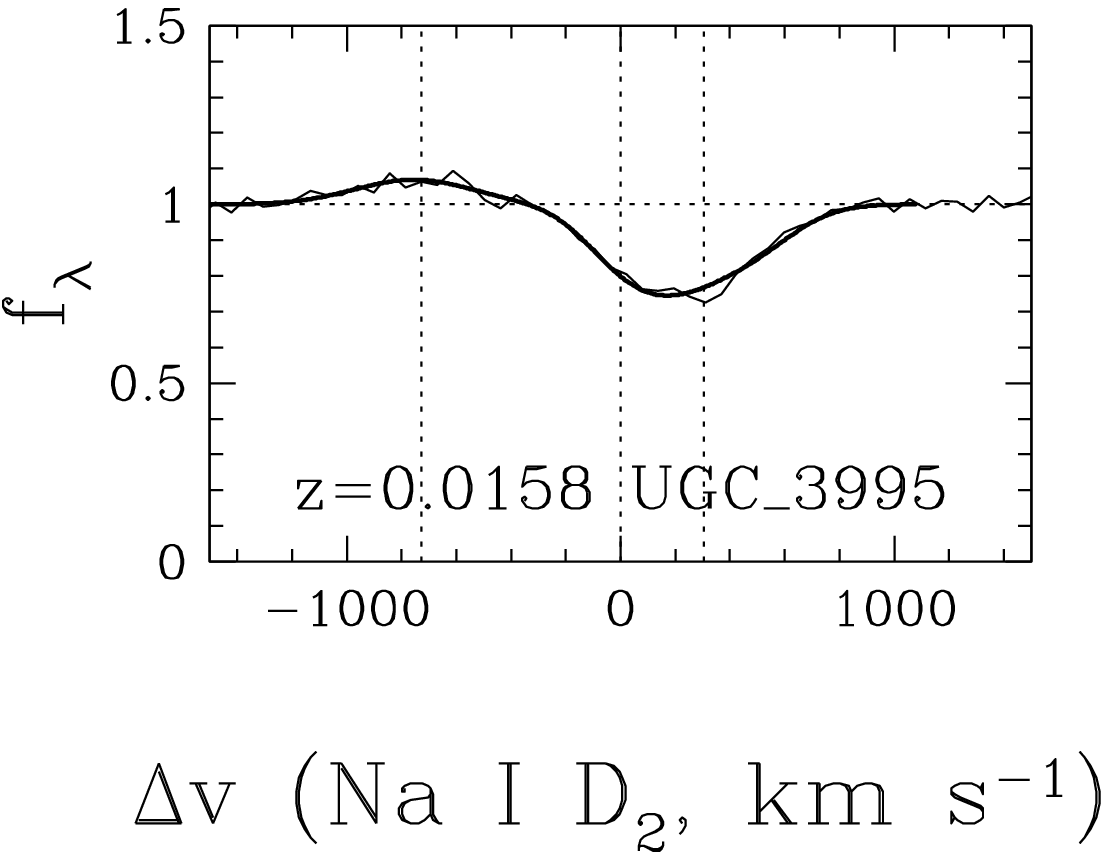}
\caption{$Continued$}
\label{sytspectraiii}
\end{figure}

\begin{figure}[t]
\epsscale{1.0}
\plotone{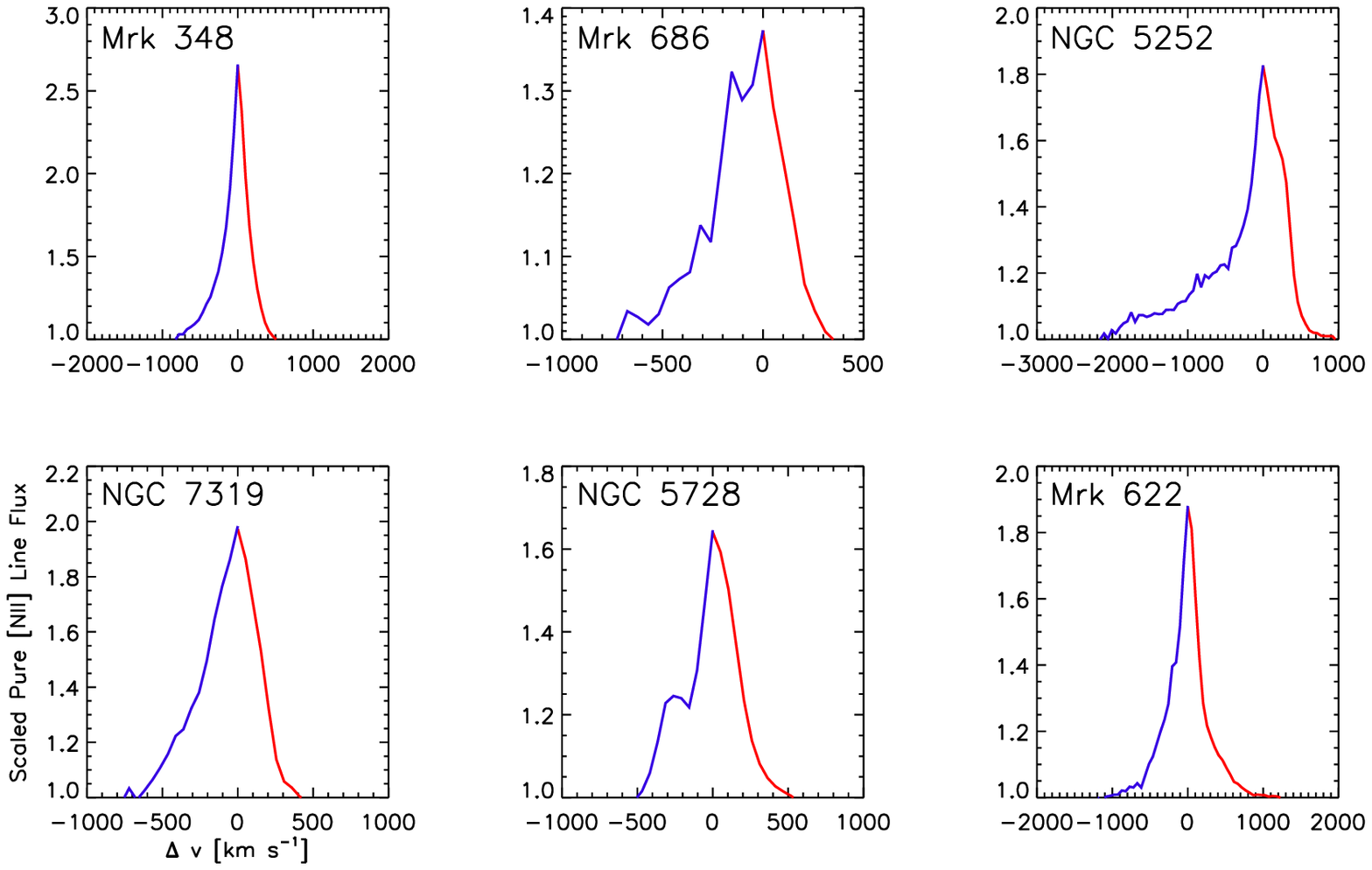}
\caption{Plots showing the \ntla~line for $\Delta v$ $<$ 0 and
  (properly scaled) \ntlb~line for $\Delta v$ $>$ 0 in 5 Seyfert 2
  galaxies suspected of having blue emission-line asymmetry.  The
  velocity scale is relative to systemic, and colors correspond to
  blue wing and red wing.  The first four objects have noticeable
  blue-wing asymmetry, and the fifth (NGC~5728) shows a prominent blue
  bump in the emission line.  The lower right panel (Mrk~622) is an
  example of an object which does not show an obvious asymmetry.}
\label{bela}
\end{figure}

\setcounter{figure}{6}

\begin{figure}[t]
\epsscale{1.0}
\plotone{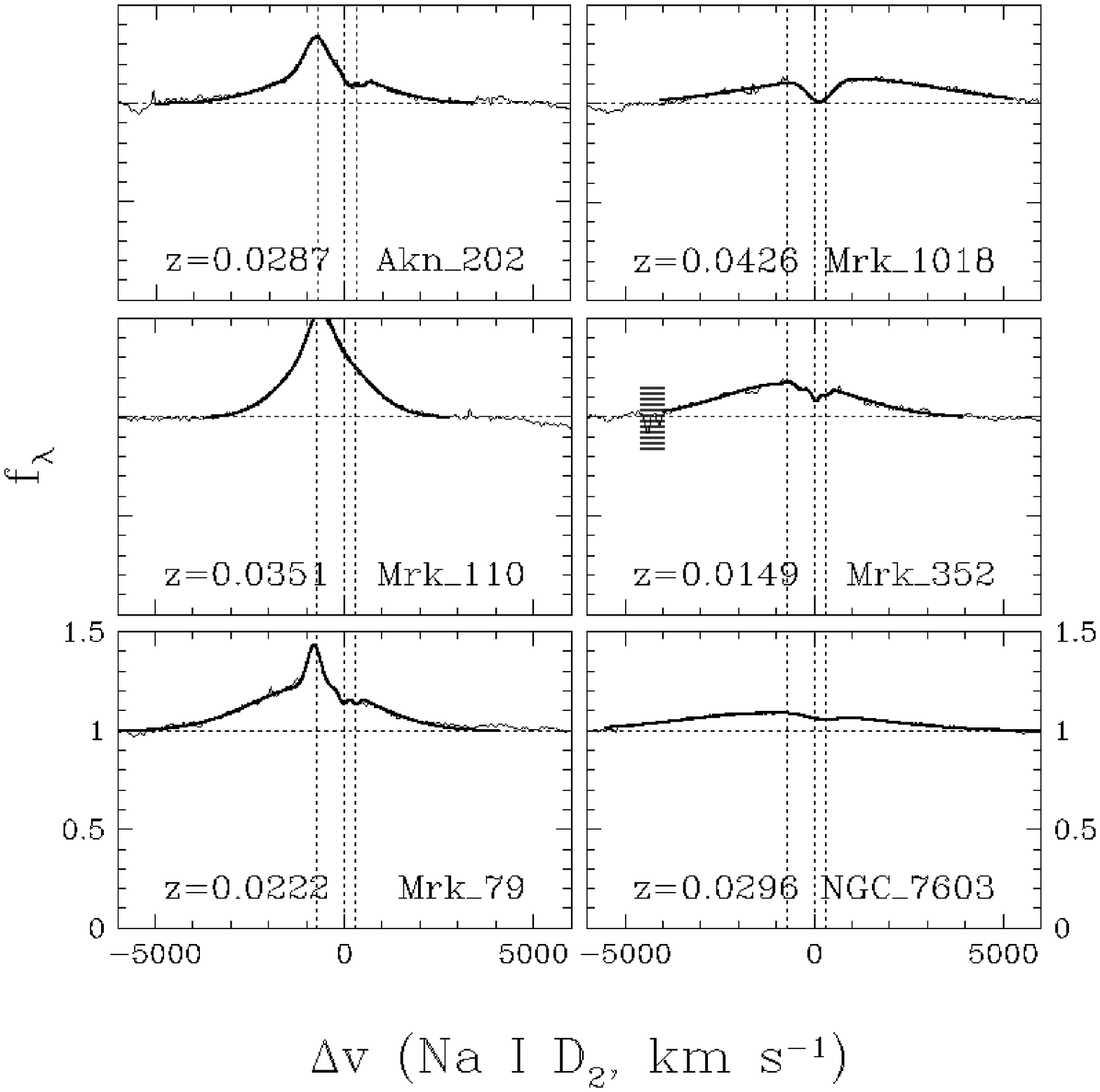}
\caption{Spectra of the \heo emission + \nad absorption region in
  infrared-faint Seyfert 1 galaxies.  See Figure \ref{sytspectrai} for
  details.  A region of horizontal lines is
  used to block out atmospheric absorption in the case of Mrk 352.}
\label{syospectrai}
\end{figure}

\setcounter{figure}{6}

\begin{figure}[t]
\epsscale{1.0}
\plotone{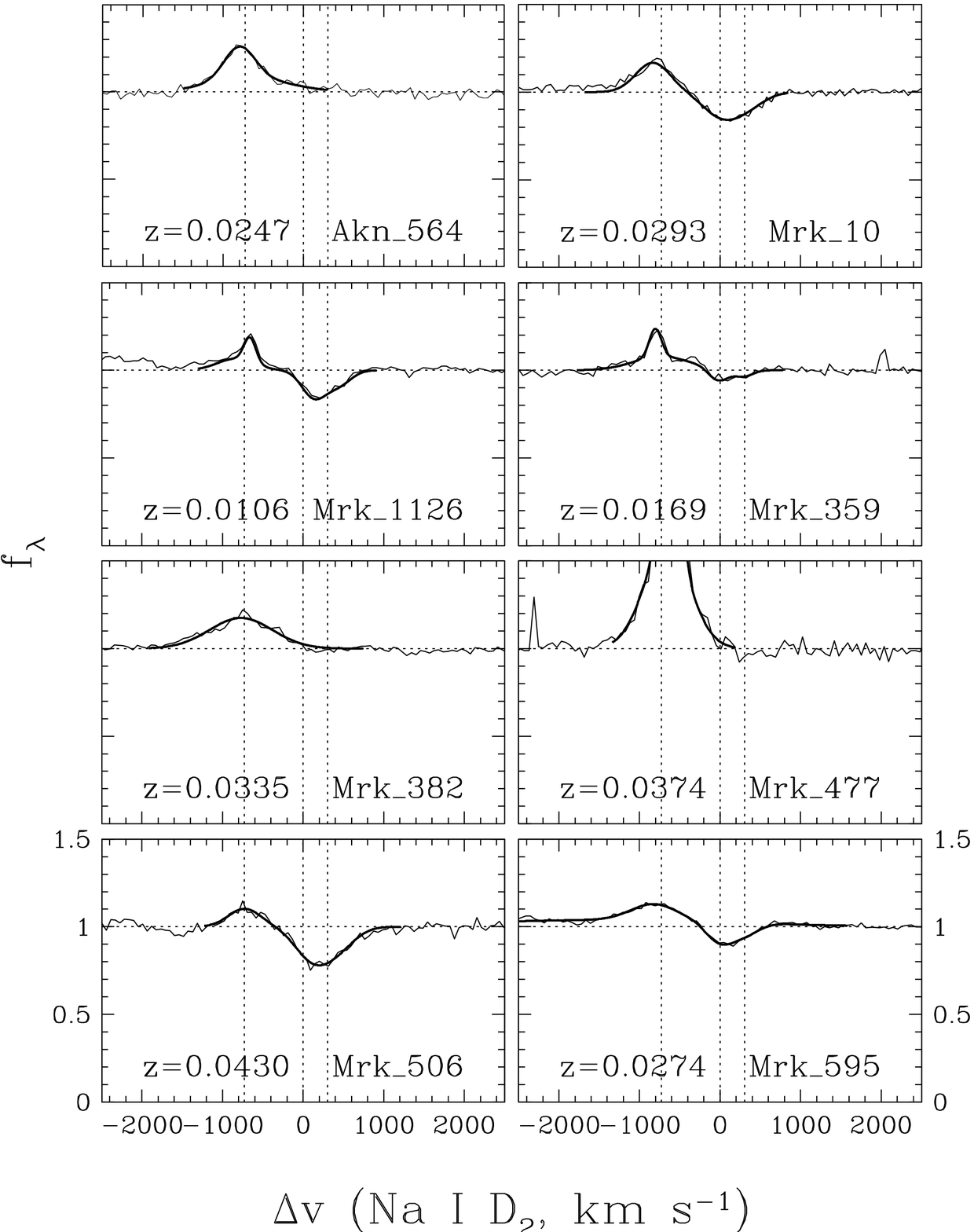}
\caption{$Continued$}
\label{syospectraii}
\end{figure}

\setcounter{figure}{6}

\begin{figure}[t]
\epsscale{1.0}
\plotone{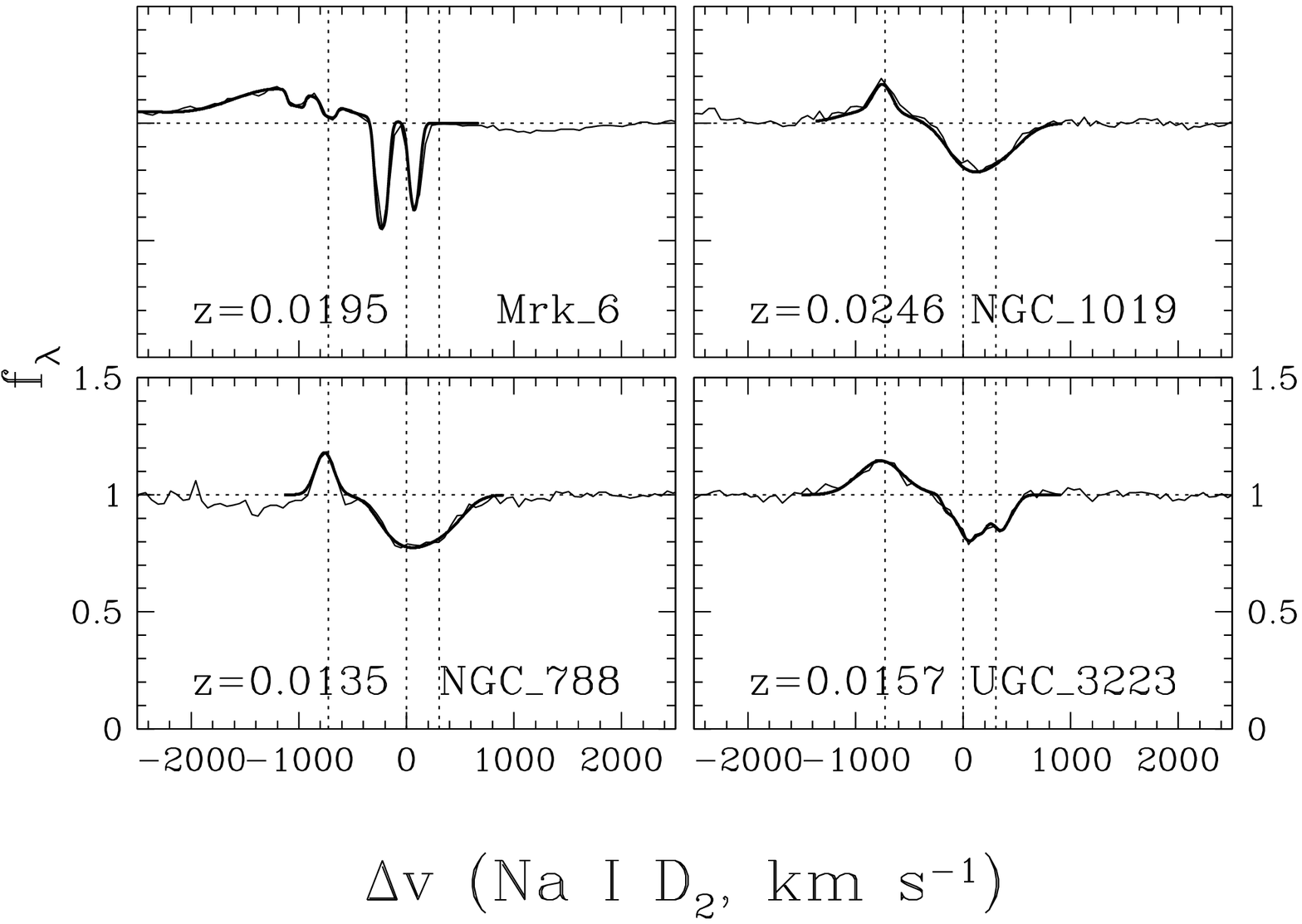}
\caption{$Continued$}
\label{syospectraiii}
\end{figure}

\begin{figure}[t]
\epsscale{1.0}
\plotone{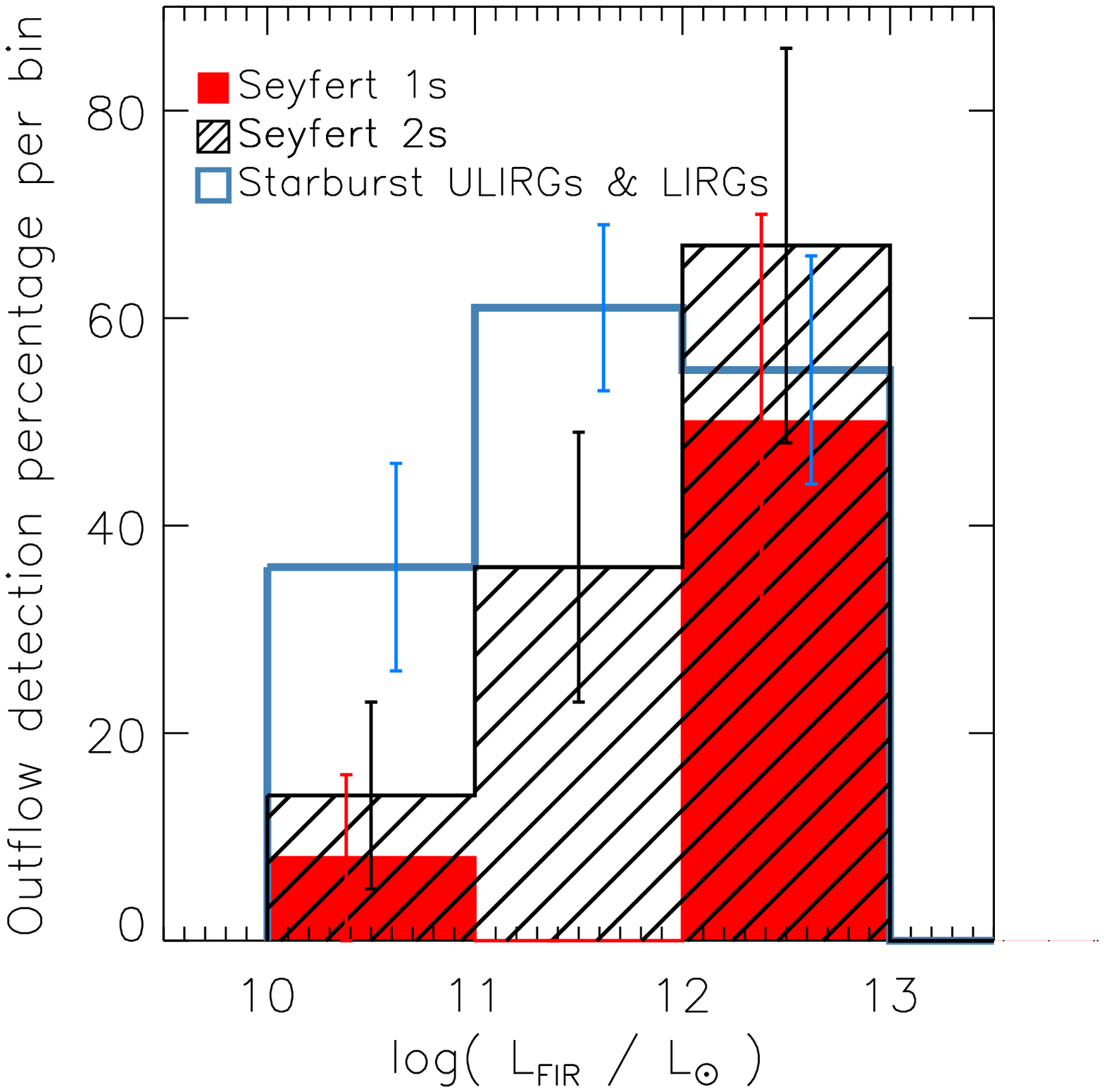}
\caption{Histogram showing the fraction of Seyfert 1s, Seyfert 2s, and
  starburst ULIRGs and LIRGs with outflows as a function of \lfir.
  The Seyfert categories include both IR-faint Seyferts from this
  study and IR-luminous Seyferts from RVS05c.  Starburst ULIRGs \&
  LIRGs are from RVS05b.  The error bars are 1$\sigma$, assuming a
  binomial distribution. Note the trend of increasing outflow
  detection rate with increasing \lfir\ for both starbursts and
  Seyferts.}
\label{detrateplots}
\end{figure}

\begin{figure}[t]
\epsscale{1.15}
\plottwo{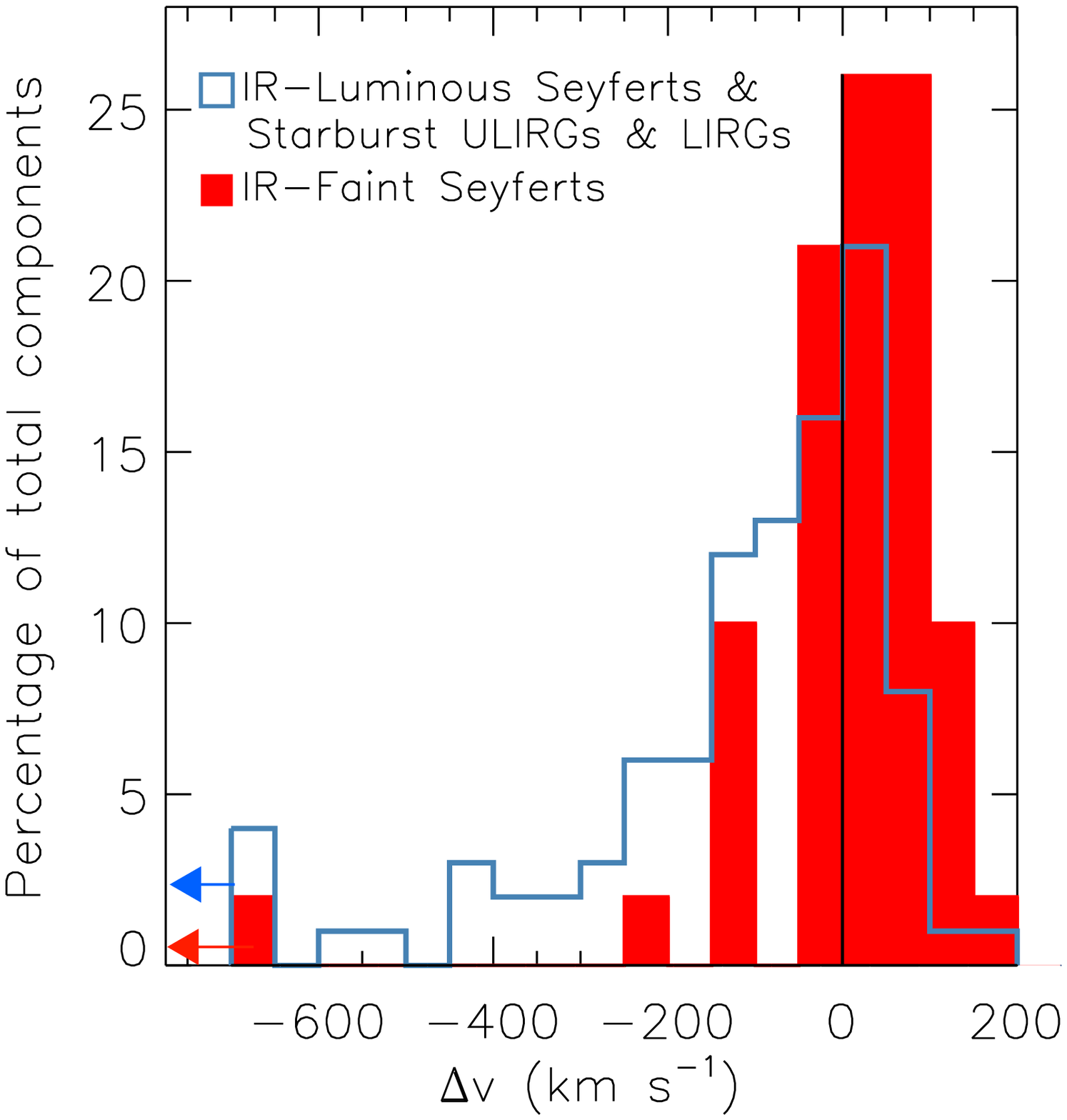}{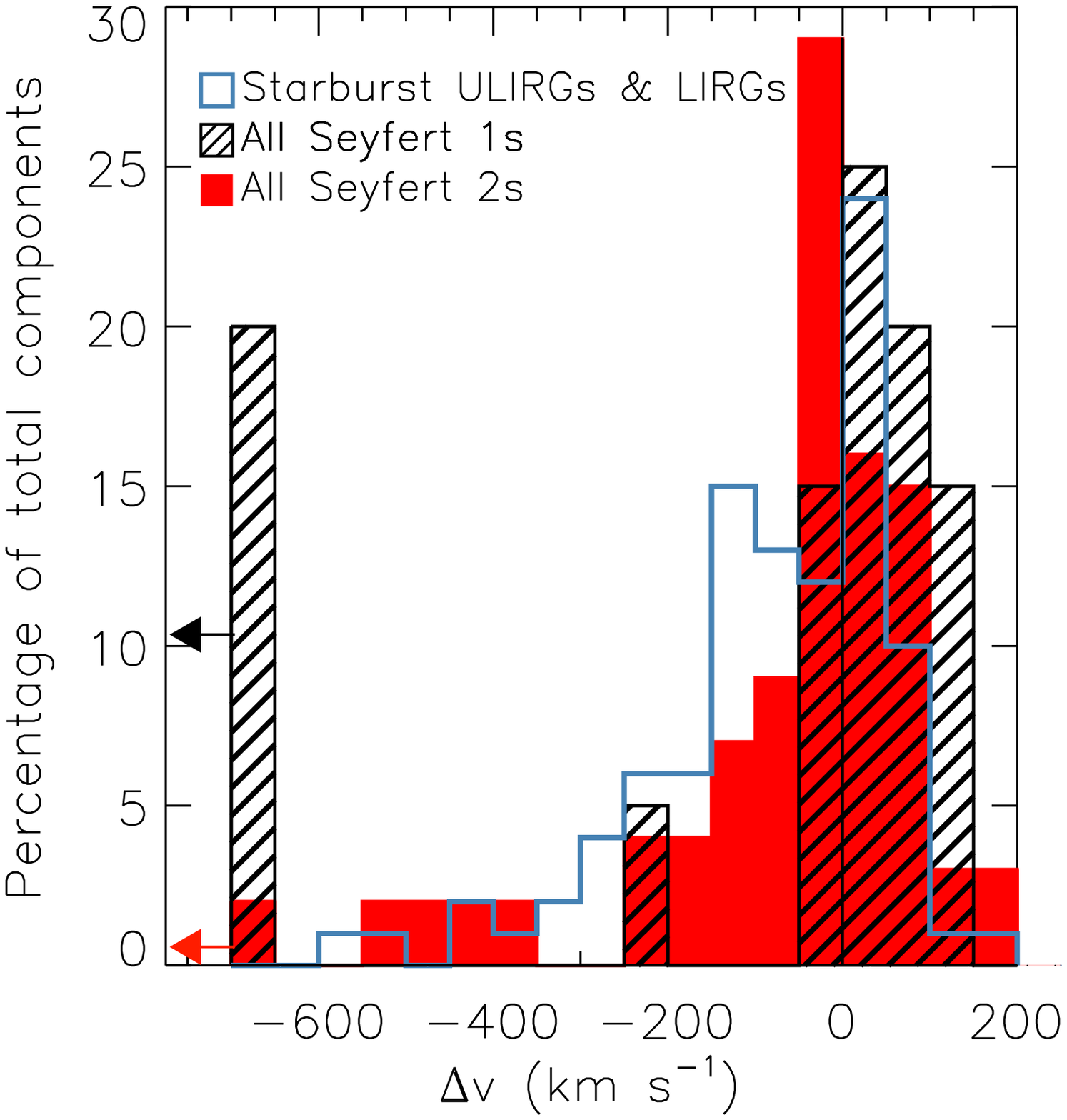}
\caption{Left: Histogram showing the percentage of velocity components
  as a function of $\Delta v$ for the IR-faint Seyferts from this
  study, the IR-luminous Seyferts from RVS05c, and the starburst
  ULIRGs \& LIRGs from RVS05b.  Objects at $\Delta v$ $=$ $-$700
  \kms~actually have $\Delta v$ $\le$ $-$700 \kms; the axis is
  truncated since most components have $\Delta v$ $>$ $-$700 \kms.
  Solid black line at $\Delta v$ $=$ 0 highlights the division between
  inflows and outflows. Note the lack (excess) of outflowing
  (infalling) components in IR-faint Seyferts relative to the other
  objects. Right: Same as left, but the IR-faint Seyferts from this
  study and IR-luminous Seyferts from RVS05c have been combined
  together and divided into Seyfert types. Note the excess of
  infalling and high-velocity outflowing components among Seyfert 1 
  galaxies.  }
\label{velhistplots}
\end{figure}

\begin{figure}[t]
\epsscale{1.1}
\begin{center}
\plottwo{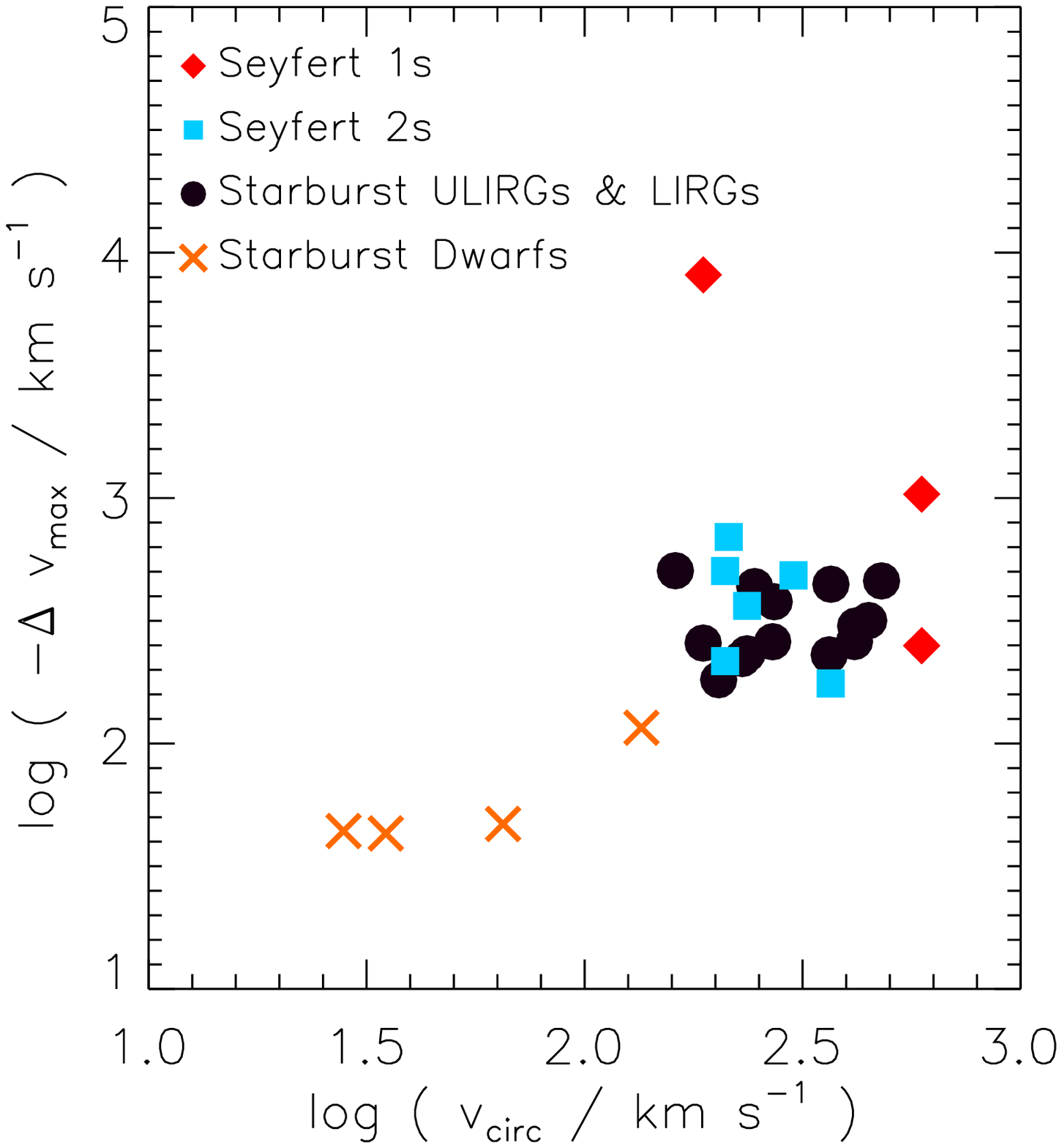}{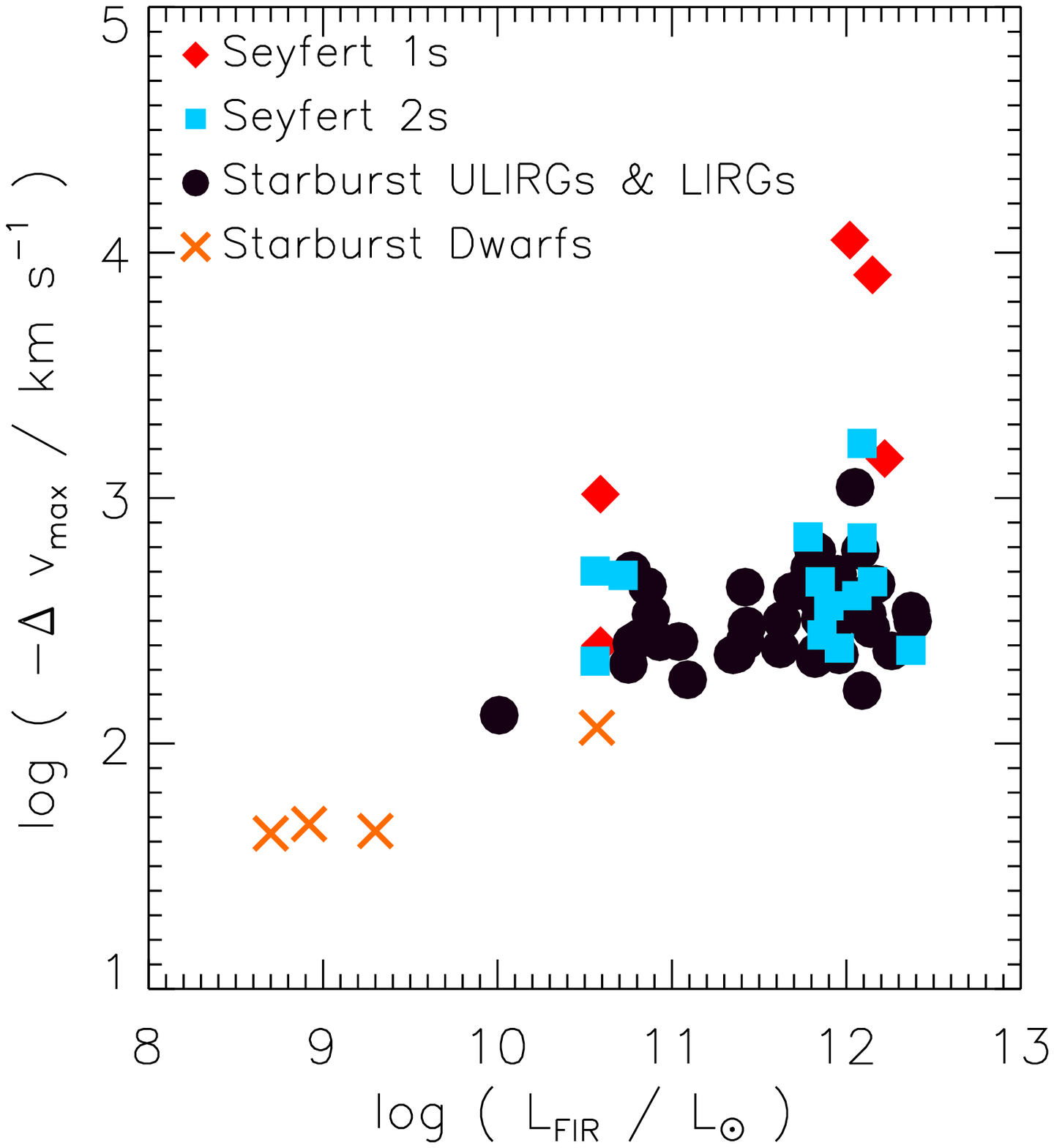}
\end{center}
\caption{Left: Maximum outflow velocity versus circular velocity.  The
  Seyfert 1 and Seyfert 2 categories include data from this study and
  from RVS05c (both IR-faint and IR-luminous Seyferts).  The starburst
  ULIRG \& LIRG category is composed of data from RVS05b.  The data on
  the starburst dwarfs are from Schwartz \& Martin (2004).  Right:
  Same as left, but for maximum outflow velocity versus \lfir.  The
  outflow kinematics in Seyfert 2s (1s) are similar to (different
  from) those of the starburst ULIRGs and LIRGs.}
\label{dvmaxplots}
\end{figure}

\begin{figure}[t]
\epsscale{1.1}
\plottwo{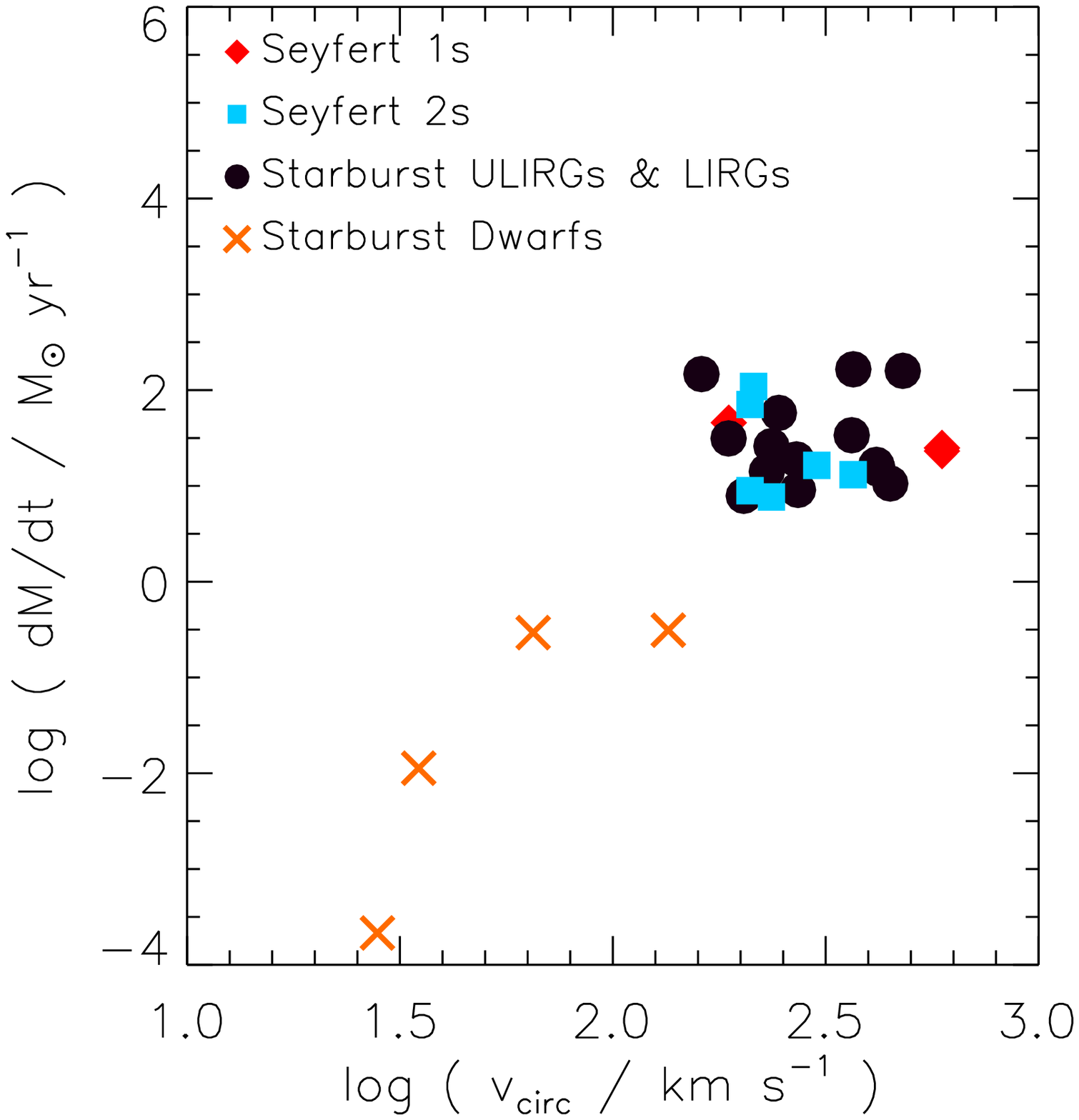}{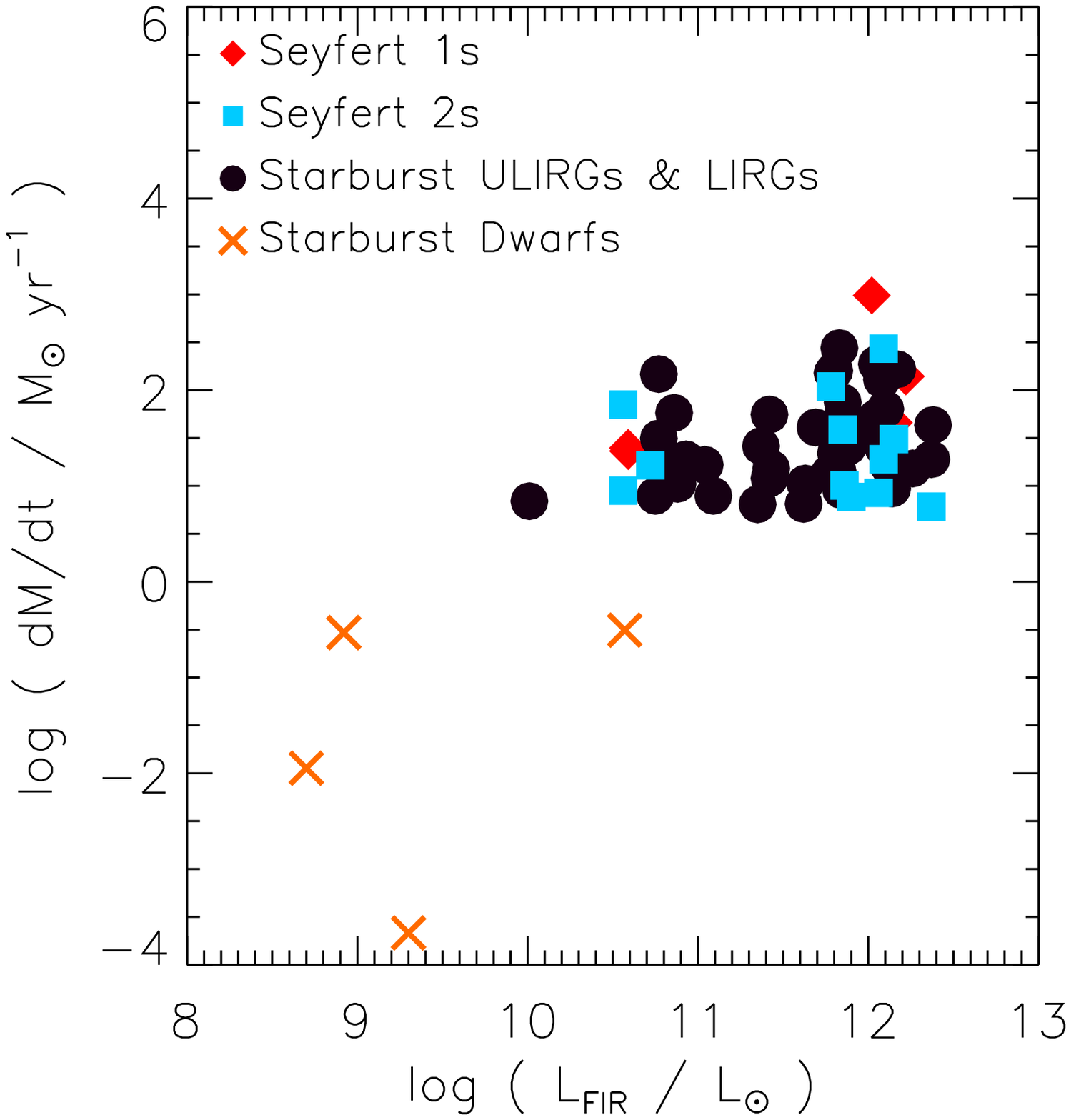}
\caption{Left: Mass outflow rate versus circular velocity. See
  Figure~\ref{dvmaxplots} for descriptions of the various categories.
  Right: Same as left, but for mass outflow rate versus \lfir.}
\label{dmdtplots}
\end{figure}

\begin{figure}[t]
\epsscale{1.1}
\plottwo{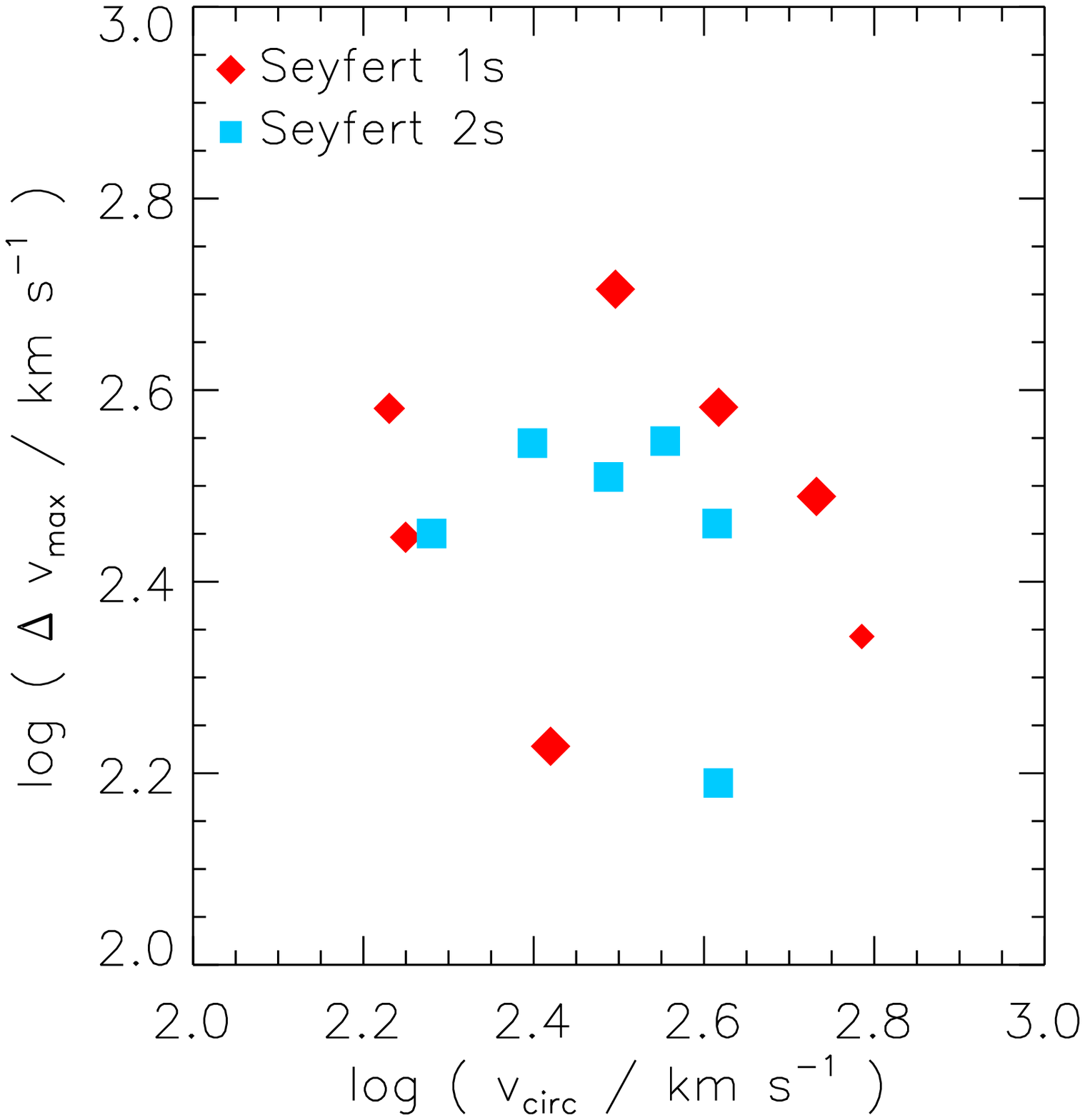}{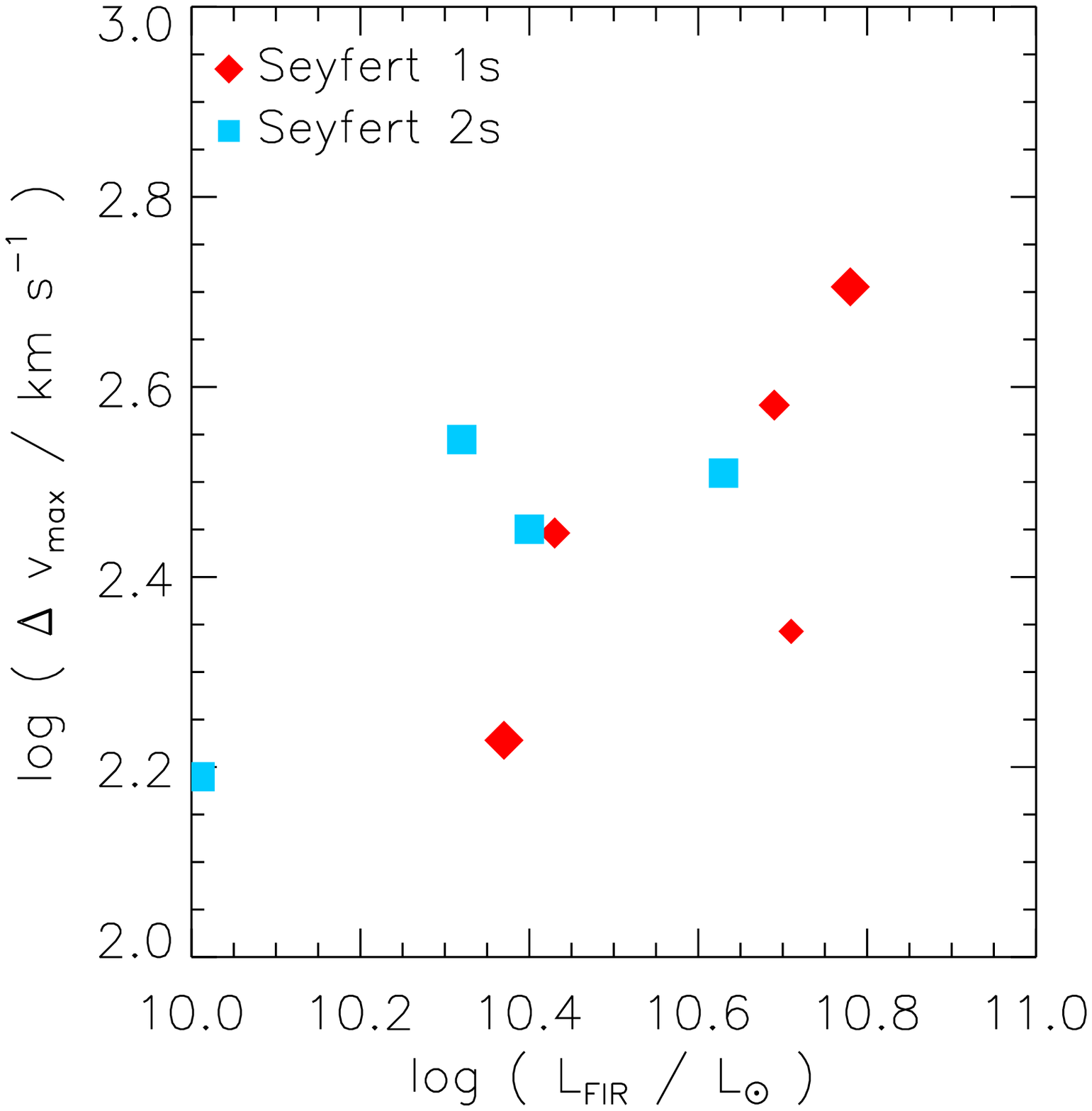}
\caption{Left: \dvmax\ versus circular velocity for inflows detected
  in IR-faint Seyfert galaxies.  Smaller points are those with
  uncertain redshifts.  Right: Same as left, but for
  \dvmax~vs. \lfir. No obvious trend is seen.}
\label{inplots}
\end{figure}

\end{document}